\documentclass[twocolumn,twocolappendix]{aastex63}
\usepackage[utf8]{inputenc}
\usepackage{placeins}
\usepackage{graphicx}
\usepackage{amsmath}

\newcommand{\inv}{$^{-1}$ }
\newcommand{\invnospace}{$^{-1}$}
\newcommand{\al}{$\alpha$}
\newcommand{\als}{$\alpha$ }
\newcommand{\fe}[1]{[#1/Fe]}

\begin{document}

\title{A Swing of the Pendulum: The Chemodynamics of the Local Stellar Halo Indicate Contributions from Several Radial Merger Events}

\author{Thomas Donlon II}
\affiliation{Department of Physics, Applied Physics and Astronomy, Rensselaer Polytechnic Institute, 110 8th St, Troy, NY 12180, USA}
\correspondingauthor{Thomas Donlon II}
\email{donlot@rpi.edu}

\author{Heidi Jo Newberg}
\affiliation{Department of Physics, Applied Physics and Astronomy, Rensselaer Polytechnic Institute, 110 8th St, Troy, NY 12180, USA}

\begin{abstract}
We find that the chemical abundances and dynamics of APOGEE and GALAH stars in the local stellar halo are inconsistent with a scenario in which the inner halo is primarily composed of debris from a single, massive, ancient merger event, as has been proposed to explain the \textit{Gaia}-Enceladus/Gaia Sausage (GSE) structure. The data contains trends of chemical composition with energy which are opposite to expectations for a single massive, ancient merger event, and multiple chemical evolution paths with distinct dynamics are present. We use a Bayesian Gaussian mixture model regression algorithm to characterize the local stellar halo, and find that the data is best fit by a model with four components. We interpret these components as the VRM, Cronus, Nereus, and Thamnos; however, Nereus and Thamnos likely represent more than one accretion event because the chemical abundance distributions of their member stars contain many peaks. Although the Cronus and Thamnos components have different dynamics, their chemical abundances suggest they may be related. We show that the distinct low and high \als halo populations from Nissen \& Schuster (2010) are explained by VRM and Cronus stars, as well as some in-situ stars. Because the local stellar halo contains multiple substructures, different popular methods of selecting GSE stars will actually select different mixtures of these substructures, which may change the apparent chemodynamic properties of the selected stars. We also find that the Splash stars in the solar region are shifted to higher $v_\phi$ and slightly lower [Fe/H] than previously reported.
\vspace{1cm}
\end{abstract}

\section{Introduction} \label{sec:intro}

It is currently our understanding that the Milky Way (MW) halo was primarily built up by accretion events \citep{Helmi2020}. Streams of stars stripped from satellites that are currently in the process of falling into the MW \citep{NewbergCarlin2016} provide evidence of this accretion history. Within the last several years, it has become apparent that the MW halo contains merger debris from at least one major merger event that is no longer identifiable in stellar density alone, but also requires velocity information of stars and/or compact objects in the inner and outer MW halo \citep{Belokurov2018,Helmi2018,Donlon2019,Donlon2020,Kruijssen2020,Naidu2020,Naidu2021,Horta2021,Malhan2022}.

The primary motivation for study of MW halo substructure is cosmological: the present-day configuration of the MW allows us to test our theories for galaxy formation and the early Universe. One of the first models that attempted to explain how the MW formed was the monolithic collapse model \citep{Eggen1962}, which proposed that large galaxies formed out of free-fall collapse of massive, uniform gas clouds.

The figurative ``pendulum'' of public thought swung the other way when \cite{Searle1978} proposed a competing hierarchical model of halo formation, as the generally-accepted standard model of $\Lambda$-CDM cosmology \citep{DekelSilk1986,Frenk1988,NFW1996} favors a stratified collapse model \citep{White1978,FrenkWhite1991}. In this model, many small pockets of star formation form out of the collapse of gas clouds. These clusters are gravitationally attracted to one another, and the accretion of lower-mass objects onto larger halos causes the build-up of progressively more massive objects over time. The popular opinion of how the MW's halo was formed transitioned away from a single collapse and towards a picture of continuous accretion. 

In order to provide constraints for cosmological models, it is useful to identify these accretion events within our Galaxy. Numerical models of halo formation were examined by \cite{Helmi1999a}, who concluded that a disrupting satellite would leave ``few obvious asymmetries [...] in the distribution of particles in configuration space,'' and that identification of halo substructure would require 3-dimensional velocity measurements. The same year, \cite{Helmi1999b} identified one such accretion event in halo stars, based on their kinematics. They estimated that the local solar region is populated by 300 to 500 low-mass stellar streams, which would cause the stellar halo to appear phase mixed even if each individual accretion event was not actually phase mixed.

Again, the pendulum swung back with the discovery that the Sagittarius spheroidal dwarf galaxy \citep[Sgr dSph,][]{Ibata1994} had a tidal stream that was apparent in the density of stars \citep{Yanny2000,Ibata2001,Newberg2002,Majewski2003}. With the realization that it was possible to detect stellar halo substructure using only the positions of stars, tidal streams became a popular way to study the halo. Presently, we know of 70+ stellar streams populating the Galactic halo (a compilation of known streams is available in the \verb!galstreams! python package, \citealt{Mateau2017}), although new streams are constantly being identified and characterized \citep[e.g.][]{Ibata2021,Martin2022}, and some streams that were previously thought to be independent may be combined into the same structure after further analysis \citep[e.g.][]{Li2021}.

At the same time, it also became apparent that there were MW halo structures that did not appear to be stellar streams, but were nonetheless apparent in stellar density. These include the Virgo Overdensity \citep[VOD,][]{Vivas2001,Newberg2002,Juric2008,Donlon2019} and the Hercules Aquila Cloud \citep[][]{Belokurov2007,Gryncewicz2021}, which were both identified as overdensities in the locations of halo stars. These structures, along with the existence of the MW halo's stellar break at $\sim$25 kpc \citep{Deason2011}, were interpreted as either the apocenter pile-up of a few accreted satellites, or a single major merger. 

This ``ancient last major merger'' hypothesis became especially popular. Decades ago, it was proposed that the MW thick disk was the result of a major merger heating the MW's proto-disk \citep{Quinn1986,Velazquez1999}. This merger event would have to be sizeable compared to the proto-MW in order to induce a sufficiently large change in the velocity dispersion of the proto-disk stars \citep[e.g.][]{Grand2020}. 

The large age of thick disk stars meant that the merger event had to have been accreted early on in the MW's history (at least 8 Gyr ago), as the end of thick disk star formation would mark the time of the merger event. It was also argued by \cite{Deason2013} that the stellar break in the MW halo could be explained by a major accretion event 6-9 Gyr ago, and \cite{Deason2015} found that the ratio of blue straggler stars to blue horizontal branch stars in the MW halo favored an accretion history consisting of a few more-massive dwarfs rather than many smaller objects. This argument was followed up by \cite{Deason2018}, who concluded that the metallicity and location of the stellar break could only be explained by a scenario involving the accretion of a single massive dwarf progenitor. The stage had been set; all that remained was to locate this ``last major merger.''

With the advent of the \textit{Gaia} survey \citep{GaiaCollaboration2016}, parallaxes and proper motions of billions of stars within several kpc of the Sun became available, which made it possible to perform large-scale kinematic and dynamical analysis on the local Solar region. Multiple groups rapidly identified the ancient last major merger of the MW. \cite{Belokurov2018} found that the local stellar halo stars predominantly had orbits with high radial anisotropy, and claimed that these very radial stars composed the debris of the ancient last major merger; they named this merger the ``Gaia Sausage''. Simultaneously, \cite{Helmi2018} identified a population of local halo stars based on their kinematics and chemical abundances, and claimed that they composed the debris of the ``ancient last major merger,'' which they named the ``\textit{Gaia}-Enceladus'' structure. This structure is typically referred to as the \textit{Gaia} Sausage/Enceladus (GSE) structure.

It was natural at this time to associate these halo stars with the ancient last major merger concept that had occupied the literature for several decades. Since the ancient last major merger was also thought to dominate the stellar halo, it was also reasonable to associate the VOD and HAC with this merger event \citep{Simion2019}. Notably, \cite{Naidu2021} were able to recover the general shape of the stellar halo with a simulation of a very massive (2.5:1) disky progenitor that fell into the MW on a retrograde orbit around 8 Gyr ago. 

In addition to the dynamical arguments for the existence of the GSE, numerous independent studies of spectroscopic data \citep{Helmi2018,Vincenzo2019,Naidu2020,Feuillet2020,Hasselquist2021,Buder2022} found that the metallicity distribution function (MDF) for GSE stars is consistent with the expected MDF of a massive dwarf galaxy \citep{Kirby2013}. Namely, the MDF of GSE stars has a single peak near [Fe/H] $\sim$ -1.2, with an extended metal-poor tail. 

However, some evidence argues against the ancient major merger scenario. It was found that the formation of a thick disk does not require a major merger event; if the MW's proto-disk was lumpy rather than uniform, it is possible to generate a disk system that is bimodal in both chemical abundances and velocity dispersion \citep{Clarke2019,Amarante2020,BeraldoeSilvia2020}. Further, \cite{BeraldoeSilva2021} showed that the thin and thick disks likely formed at the same time early on in the MW's history. These studies call into question the requirement that the radial stars in the local solar region must have been accreted at early times in order to puff up the thick disk. 

The inner-halo structure known as the ``Splash'' \citep{Belokurov2020} initially supported an ancient last major merger scenario, because it was thought to originate from heating of the inner MW proto-disk stars to orbits with large vertical heights, probably with the same merger that heated the thick disk. However, this structure also can be explained if the early MW contained a lumpy proto-disk \citep{Amarante2020}, or instead could have formed within gas outflows that were thrown out from the disk and then fell back into the inner MW \citep{Yu2020}.

The morphology of these radial stars in phase space also called the large age of the structure into question. \cite{Donlon2019} found that the VOD could be recreated using an $N$-body simulation of a dwarf galaxy on a radial orbit that began disrupting 2 Gyr ago, but they were not able to create overdensities in the halo with long integration times. They named the merger event that was responsible for creating the VOD the ``Virgo Radial Merger'' (VRM).  Following those results, \cite{Donlon2020} found that the VOD and HAC regions contained shell structure, which is an indicator of a radial merger event. They found that if the progenitor of the VOD and HAC had been disrupted more than 5 Gyr ago, then we would not expect to observe any shell structure in the MW. Based on this phase mixing argument, as well as reverse orbit integration of the shell structures, \cite{Donlon2020} argued that the progenitor of the VOD and HAC collided with the Galactic Center within the last 3 Gyr (although \citealp{Donlon2020} used a spherical halo potential for their orbit integrations, which might artificially decrease the amount of time for which large-scale asymmetries remain visible based on the findings of \citealp{Han2022}). Additionally, \cite{Grunblatt2021} computed asteroseismic ages for several GSE stars, and found that some stars had been formed as recently as 6 Gyr ago; this called the proposed infall time of the GSE progenitor into question, but allowed for a more recent merger time. 

The recent time of the VRM appeared to be in tension with other published results. First, the thick disk could not have been heated 3 Gyr ago. That is okay if the thick disk was heated by another mechanism; the more recent merger would interact with a more massive MW, and should not cause a recent disk heating event \citep[see the mass/velocity scales of ][]{Grand2020}. The VRM could instead have produced the observed increases in star formation rate in the MW disk around 2.5-2.7 Gyr ago \citep{Snaith2014,Mor2019}, although these starbursts could instead be caused by the passage of the Sagittarius Dwarf Galaxy \citep[Sgr dSph,][]{Ruiz-Lara2020}. One might worry about a recent merger time given that the GSE/VRM stars are predominantly old; however, the peak ages of these stars are not necessarily tied to the time that their progenitor was accreted. The best example of this is that the Sgr dSph stars appear to have been formed before 8 Gyr ago \citep{deBoer2015,Hasselquist2021}, even though the dwarf is still bound today. For a detailed discussion of these points, we direct the reader to Section 7 of \cite{Donlon2020}.

Since the initial discoveries of the ancient last major merger, there have been numerous works that propose the existence of different phase-mixed merger events within the MW halo. These include the strongly retrograde ``Thamnos'' structure \citep{Koppelman2019b}, the ``Kraken'' structure that was predicted via populations of globular clusters \citep{Kruijssen2020}, the ``Inner Galaxy Structure'' (also called ``Heracles'') that was identified in halo stars \citep{Horta2021}, multiple retrograde structures identified in halo stars, \citep{Naidu2020}, and the ``Pontus'' structure identified in integral clustering of MW globular clusters and dwarf galaxies \citep{Malhan2022}. These different halo structures all make up the phase-mixed stellar halo, and it is not likely that they all come from a single massive merger event. However, none of these structures are thought to dominate the stellar halo near the location of the Sun.

Of particular importance to this work are the halo structures identified in the solar neighborhood by \cite{Donlon2022}. While previous work called into question the arguments for an ancient major merger or its association with halo structures like the Virgo Overdensity, \cite{Donlon2022} found multiple radial structures in a volume of halo stars that should have been dominated by only the GSE.

Using local dwarf stars identified by \cite{KimLepine2021} combined with proper motions from \textit{Gaia}-EDR3, \cite{Donlon2022} identified VRM stars using the criteria that they are relatively metal-rich, are slightly counter-rotating from the disk, and have large radial velocities as viewed from the Galactic Center ($v_R$). In addition, \cite{Donlon2022} identified two new structures in these dwarf stars. The first, named ``Cronus,''  was identified as a relatively metal rich structure with low $V_R$, and is slightly co-rotating with the MW disk. The second, named ``Nereus,'' was identified as a metal-poor structure with no net rotational velocity, and has $v_R$ somewhere between those of Cronus and the VRM. Notably, all of these structures are made up of stars that would typically be classified as belonging to the GSE structure. \cite{Donlon2022} conclude that the combination of the VRM, Nereus, and Cronus structures cannot be explained by a single ancient, massive merger event, and that the structure commonly referred to as the GSE is really a combination of all of these components. 

In this work, we classify local Solar neighborhood halo stars using their chemical abundances as well as their dynamical properties. We find that these local halo stars cluster into four components of the stellar halo, which correspond to the VRM, Nereus and Cronus components of the halo that were identified in \cite{Donlon2022}, as well as Thamnos. We again conclude that the inner halo stars that are commonly attributed to the GSE merger event actually belong to a combination of these four halo components. Our results suggest the VRM was only recently accreted by the MW, Cronus was accreted early on in the MW's history, and Nereus and Thamnos are made up of multiple smaller radial merger events that may have a range of accretion times.

We believe that we are experiencing yet another shift in our understanding of the MW's inner stellar halo. At first it was thought that galaxies formed under monolithic collapse, until hierarchical collapse was shown to describe the Universe. In the past, the literature was certain that stellar streams would only be found kinematically -- until a stream was discovered using the positions of stars. The popular understanding of the MW's inner stellar halo has transitioned from a belief that the local stellar halo was built up from 300-500 structures, to a belief that the majority of the MW's inner stellar halo was deposited by a single merger event. Now, with substantial evidence that the local stellar halo was actually built up from several merger events, we find that the pendulum is once again swinging, and that our understanding of the stellar halo must now allow for several mergers contributing to the accreted halo stars in the local Solar neighborhood.  

\section{Data} \label{sec:data}

\begin{figure}
    \centering
    \includegraphics[width=\linewidth]{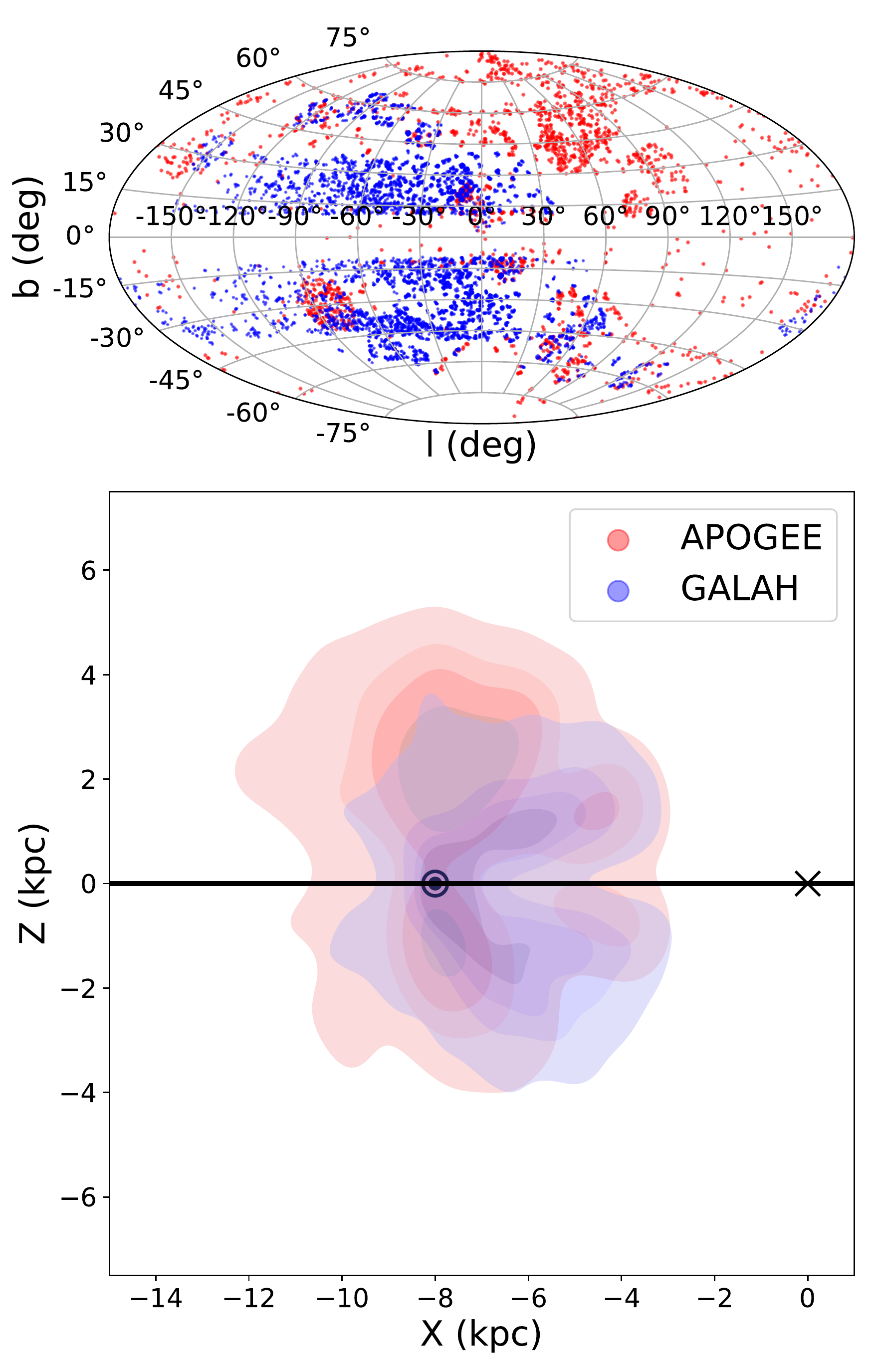}
    \caption{Spatial distribution of stars in the dataset. Stars from APOGEE are shown in red, and stars from GALAH are shown in blue. The top panel shows the distribution of the data on the sky in Galactic coordinates. The bottom panel shows the distribution of the data in Galactic Cartesian coordinates, where the circular symbol is the position of the Sun, the ``x'' symbol is the position of the Galactic center, and the solid black line is the Galactic midplane. Because the telescopes for these surveys are located in different hemispheres, the two surveys probe very different volumes; GALAH stars are located preferentially towards the Galactic center, while APOGEE stars are mostly in the north Galactic cap.}
    \label{fig:data_sky}
\end{figure}

In this section, we describe the process of obtaining and cleaning the data that we analyze in this work.

We constructed a dataset consisting of stars from SDSS Data Release 17 \citep{SDSSIV,SDSSDR17} that were in the APOGEE survey \citep{SDSSAPOGEE2017}, and stars from GALAH DR3 \citep{GALAHDR3,GALAHPIPELINE}. The data from these surveys were supplemented by \textit{Gaia} EDR3 data \citep{GAIAMISSION,GAIAEDR3}.

For both of the APOGEE and GALAH samples, the six-dimensional phase space information for each star was derived using astrometric positions, distances obtained from \textit{Gaia} EDR3 parallaxes, \textit{Gaia} EDR3 proper motions, and line-of-sight velocities from the corresponding spectroscopic survey. 

Stars from both catalogs were required to satisfy the \textit{Gaia} EDR3 astrometric cuts \verb!astrometric_excess_noise! $< 0.25$, which removes stars that contain disagreements between observations of a source and the \textit{Gaia} astrometric model \citep{Lindegren2012}, and \verb!phot_bp_rp_excess_factor!  $< 1.5$, which removes stars that have contamination issues in the \textit{Gaia} BP and RP photometry \citep{Riello2018}. 

The \textit{Gaia} EDR3 parallaxes were adjusted using the \verb!gaiadr3_zeropoint! python package \citep{Lindegren2021}. Additionally, it was required that the parallax for each star be positive before zero-point offset correction, and that the relative error in the parallax was less than 15\%. We restricted our data to be within 5 kpc of the Sun, as distances derived from \textit{Gaia} parallaxes beyond this range can become unreliable.

This work uses a right-handed Galactocentric Cartesian coordinate system. The Sun is located at (-8.21, 0, 0.25) kpc \citep[consistent with][]{GravityCollaboration2019,Juric2008}, where positive $X$ points from the Sun towards the Galactic center, positive $Y$ points in the direction of the Sun's motion, and $\hat{Z} = \hat{X} \times \hat{Y}.$ The circular velocity at the Sun's position is 233.1 km s\invnospace, and the Sun's motion with respect to the LSR is (11.1, 15.17, 7.25) km s\inv (\citealp{Schonrich2010}, except $v_y$ is slightly larger than their reported value; see \citealp{Buder2021} for details). This kinematic frame was used for the GALAH DR3 value-added dynamical quantity catalog \verb!GALAH_DR3_VAC_dynamics_v2!, and we choose to use it for the APOGEE data as well for consistency. However, it should be noted that this choice of frame can have a substantial effect on the measured dynamical quantities of stars, particularly $L_z$ (see Appendix \ref{app:lz_diff}). We take the sign of the azimuthal velocity $v_\phi$ to be positive in the direction of disk motion. Note, however, that the sign of $L_z$ is negative for prograde motion.

The \verb!GALAH_DR3_VAC_dynamics_v2! catalog includes several orbital parameters, including the maximum orbital height above the plane $z_\textrm{max}$, maximum ($r_{apo}$) and minimum ($r_{peri}$) distances from the Galactic center, and orbital eccentricity $e$. It also includes energies and actions that are computed in the axisymmetric \verb!McMillan2017! potential \citep{McMillan2017} using the \verb!galpy! package \citep{Bovy2015}. These actions were computed using a Staekel fudge algorithm with a focus of 0.45. We took these dynamical values directly from the \verb!GALAH_DR3_VAC_dynamics_v2! catalog for the GALAH stars, except we negated the GALAH values of $L_z$ ($= J_\phi$) to match the definition of our coordinate system. Dynamical quantities were calculated for the APOGEE stars using the \verb!actionAngleStaeckel! class in \verb!galpy!, with the same settings as the GALAH DR3 dynamical quantity calculations \citep{Buder2021}.

We queried all available chemical abundance data for each star, which included 23 abundances for APOGEE stars and 35 abundances for GALAH stars. However, for some abundances there was only data for a few or no stars in our final dataset, so they could not be included in our analysis. For APOGEE stars, the [P/Fe] and [Cu/Fe] abundances were not included (for a total of 21 APOGEE usable abundances); for GALAH stars, the \fe{C}, \fe{ScII}, \fe{CrII}, \fe{Sr}, \fe{Rb}, \fe{Mo}, and \fe{Ru} abundances were not included (for a total of 28 usable GALAH abundances). 

Of particular interest are stellar [Mg/Mn] and \fe{Na} abundances. [Mg/Mn] is often preferred to \fe{Mg} in other works utilizing GALAH abundances \citep[e.g. ][]{Hawkins2015,Das2020,Buder2022}, due to the fact that Mg and Mn probe orthogonal nucleosynthesis processes \citep{Ting2012}. \cite{Buder2022} provide detailed reasoning for using [Mg/Mn] and \fe{Na} in their analysis, and found success with using those abundances to identify accreted halo stars.

In order to ensure that our dataset did not contain stars that are members of compact structures, we removed all stars that were located within 0.3$^\circ$ of any cluster that is listed in the \textit{Catalog of Parameters for Milky Way Globular Clusters} \citep{Harris1996} and is located within 10 kpc of the Sun. In the case of the Omega Centauri globular cluster, all stars within 0.5$^\circ$ of the center of the cluster were removed instead, due to the large apparent size of the object. 

Finally, in order to remove the majority of disk stars from our dataset, we required that all stars had velocities that differed by at least 180 km s\inv from the LSR. After this cut there were still many thick disk stars present in the data, which were removed via chemical abundance cuts (Section \ref{sec:disk_splash}). 

Additional quality cuts were made for both APOGEE stars and GALAH stars, which are outlined in the following subsections. Figure \ref{fig:data_sky} shows the spatial distribution of the cleaned dataset. Stars from the two surveys are located in different volumes due to the on-sky and distance distributions of stars in each survey.

\subsection{APOGEE Data Selection \& Cleaning}

We selected all stars from APOGEE Data Release 17 that had been processed by the ASPCAP pipeline \citep{ASPCAP}. We generally followed the cleaning procedure of \cite{Das2020}: we required that each star have \verb!STARFLAG=0! and \verb!ASPCAPFLAG=0!, which are nonzero if the APOGEE and ASPCAP pipelines experience problems while reducing the data for a star; we require that the [Fe/H], [\al/M], [Mg/Fe], [C/Fe], and [Al/Fe] abundances for these stars  are known, and that their errors are smaller than 0.15 dex; and we require that the effective temperature $T_{\textrm{eff}} > 4000\textrm{ K}$ and surface gravity $\log g > 0.5$ for each star, as APOGEE derived values are not reliable outside these ranges. We use the calibrated abundances provided by APOGEE, which are only available for giant stars.

\cite{Das2020} also required that the [N/H] and [Mn/Fe] abundances are known and have small errors. However, those additional cuts greatly reduced the number of available stars in our dataset. Since [N/H] and [Mn/Fe] are not used to remove disk/Splash contamination (Section \ref{sec:disk_splash}) or to fit a model to the APOGEE data (Section \ref{sec:model_fit}), we do not impose these restrictions on the data.

After all cuts, there were a total of 2,301 APOGEE stars in our dataset, which made up 28.3\% of our full dataset. 

\subsection{GALAH Data Selection \& Cleaning}

We selected all stars from GALAH DR3 in the \verb!GALAH_DR3_main_allstar_v2! catalog, which contains the spectroscopically derived quantities available in GALAH DR3. We then cross-matched this catalog with the \verb!GALAH_DR3_VAC_dynamics_v2! catalog, which contains dynamical information for GALAH stars.

We required that the stars have \verb!flag_sp! $ = 0$, \verb!flag_fe_h! $ =0$, and that the corresponding flags for [\al/Fe], [Mg/Fe], [Mn/Fe], and [Na/Fe] were not set. This ensured that the measurements of these abundances were of high quality. We did not remove stars that had flags on other abundance measurements, but flagged abundances were not used for analysis. Following the procedure of \cite{Buder2022}, we also required that \verb!survey! $\neq$ ``other''.

After all cuts, there were a total of 5,828 GALAH stars in our dataset, which made up 71.7\% of our full dataset. 

\section{Thick Disk \& Splash Stars} \label{sec:disk_splash}

\begin{figure*}
    \centering
    \includegraphics[width=\linewidth]{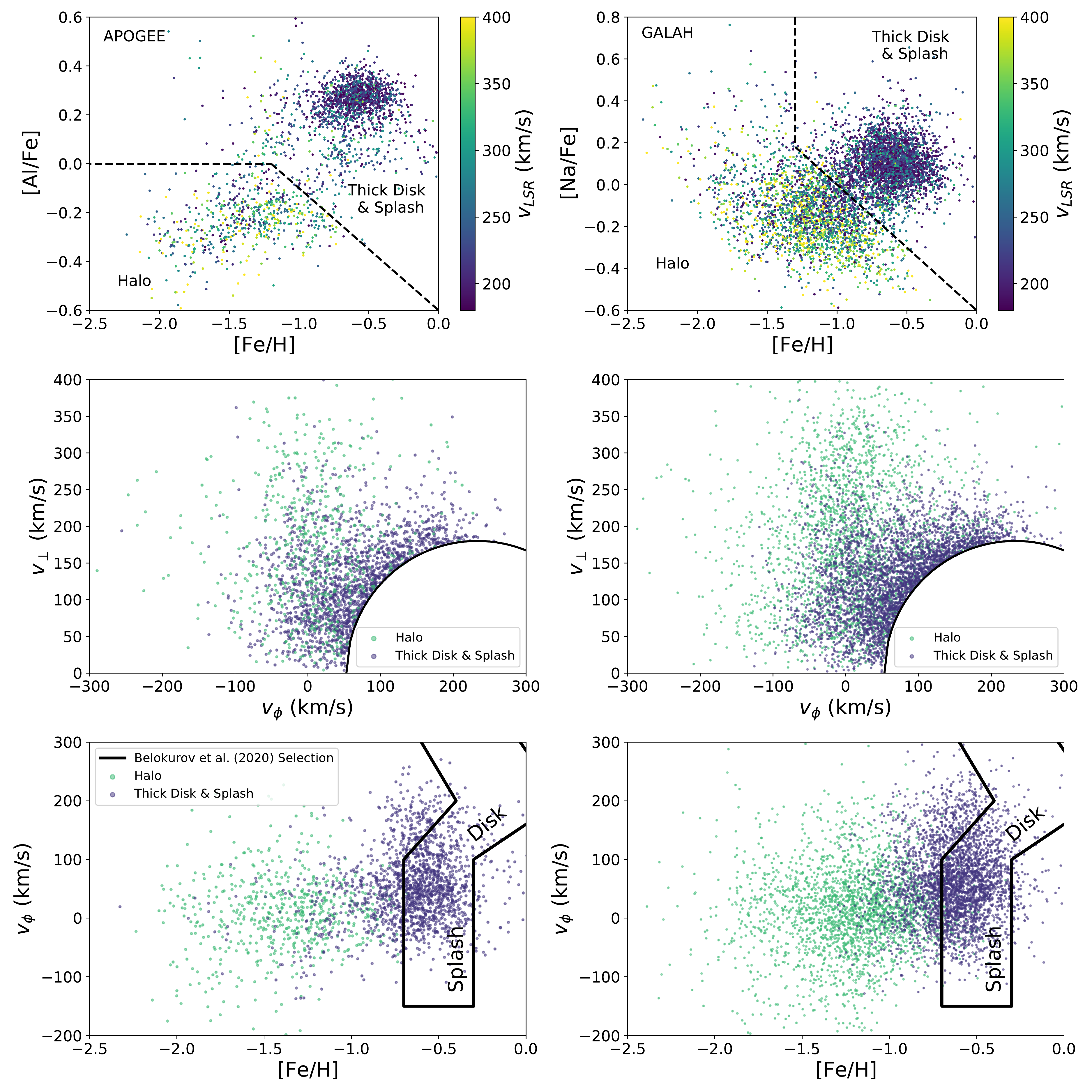}
    \caption{Chemical abundance selections of thick disk/Splash stars vs. halo stars, and their corresponding kinematics. The left column shows APOGEE data, and the right column shows GALAH data. The top row shows chemical selections of thick disk/Splash stars (dashed black lines), which separate from halo stars in abundances, as shown by their relative velocity with respect to the LSR. The middle row contains a Toomre diagram of the data split up into halo stars (green) and thick disk/Splash stars (purple). The thick disk/Splash stars have velocities that are preferentially closer to the LSR, and hug the velocity cut (solid black line). The bottom row shows a plot of $v_\phi$ vs. [Fe/H], split up into halo stars and thick disk/Splash stars. The original extent of Splash stars from \cite{Belokurov2020} is drawn over the data in black. Compared to the original \cite{Belokurov2020} selection, our thick disk/Splash selection extends to lower [Fe/H] values, does not appear to bend at large $v_\phi$ values, and does not extend down as far to negative $v_\phi$ values. The lack of a bend in the thick disk/Splash star data is because of the velocity cuts that we used to remove disk stars from the data. }
    \label{fig:disk_splash}
\end{figure*}

This section contains our procedure for selecting thick disk and Splash stars in our dataset based on their kinematics and chemical abundances, as well as a brief discussion about the properties of these stars in our data compared to other literature. This selection of thick disk \& Splash stars will be removed from our sample in order to analyze the properties of the accreted stellar halo.

The kinematic cuts made in the previous section removed the majority of thin disk stars, which have velocities very similar to the LSR. However, thick disk stars can have velocities that are different enough from the LSR that they are not removed by our kinematic cut. Additionally, \cite{Belokurov2020} showed that there was a population of metal-rich inner halo stars that appears to smoothly extend stars with thick disk [Fe/H] values down to somewhat retrograde $v_\phi$ values, which they called the ``Splash''; these stars will similarly not be removed from our velocity cut.

\subsection{Selecting Thick Disk/Splash Stars}

\cite{Das2020} showed that in APOGEE data, disk stars appeared to separate from halo stars in plots of [Al/Fe] vs. [Fe/H]. We plot our APOGEE data in [Al/Fe] vs. [Fe/H] in the top row of Figure \ref{fig:disk_splash}, except we color each star by its velocity with respect to the LSR: \begin{equation}
    V_{LSR} = \sqrt{v_R^2 + (v_\phi - 233.1)^2 + v_z^2}.
\end{equation} Thick disk/Splash stars should have smaller values of $V_{LSR}$ than halo stars that are not associated with the disk. There is a clear clump of stars with small $V_{LSR}$ at ([Fe/H], [Al/Fe]) = (-0.6, 0.3) in the APOGEE data. Additionally, there are two trails of stars leading from the low-$V_{LSR}$ clump down and to the left; these are high-\als disk stars \citep{Belokurov2022}, which we also want to remove from our dataset. These in-situ stars were selected using the criteria \begin{equation}\begin{aligned}
    \textrm{[Al/Fe]} > & -0.5\;[\textrm{Fe/H}] - 0.6, \textrm{ and} \\
    \textrm{[Al/Fe]} > & \;0.
\end{aligned}\end{equation} This cut is shown with dashed black lines in the top left panel of Figure \ref{fig:disk_splash}. 1,686 stars (73.3\%) of the cleaned APOGEE sample were classified as ``in-situ'' stars by this cut, and the remaining 615 (26.7\%) APOGEE stars were classified as accreted halo stars. 

\cite{Das2020} also used [Mg/Mn], \fe{C}, and \fe{N} to identify stars that belonged to the thick and thin disk. We chose to avoid requiring our APOGEE stars to have low-uncertainty \fe{Mn} and \fe{N} information because using those abundances greatly restricts the number of viable halo stars. While most of our APOGEE halo stars have high-quality \fe{C} abundances, we were not able to separate disk and halo stars using these \fe{C} abundances.

Chemical cuts were also used in the GALAH stars to isolate thick disk/Splash stars. However, we chose to use \fe{Na} rather than \fe{Al} for this task because there are 5,828 \fe{Na} abundances in our dataset, but only 3,937 GALAH \fe{Al} abundances. \cite{Buder2022} also chose to use \fe{Na} to determine candidate GSE stars in GALAH data instead of \fe{Al} for the same reason. There is a clump of stars with small $V_{LSR}$ at ([Fe/H], [Na/Fe]) = (-0.6, 0.15) in the GALAH data. Our selection of thick disk/Splash stars in the GALAH data is \begin{equation}\label{eq:galah_disk_cut}\begin{aligned}
    \textrm{[Na/Fe]} > & -0.6\;\textrm{[Fe/H]} - 0.6, \textrm{ and} \\
    \textrm{[Fe/H]} > & -1.
\end{aligned}\end{equation} This cut is shown as dashed black lines in the top right panel of Figure \ref{fig:disk_splash}. 3,413 stars (58.6\%) of the cleaned GALAH sample were classified as ``in-situ'' stars by this cut, and the remaining 2,415 (41.4\%) cleaned GALAH stars were classified as accreted halo stars. 

The ratio of ``in-situ'' stars to accreted stars is substantially different for the APOGEE and GALAH samples. This may be due in part to our explicit removal of the \al-rich disk stars in APOGEE data, which we were not able to do in the GALAH data. However, the main difference is likely the selection functions of the surveys: APOGEE is designed to obtain spectra of any red (primarily giant) stars, while GALAH contains a larger fraction of accreted stars because its targets were chosen with Galactic archaeology in mind. 

\subsection{Properties of Thick Disk/Splash Stars}

The middle row of Figure \ref{fig:disk_splash} shows $v_\perp$ vs. $v_\phi$ for halo stars (green) and thick disk/Splash stars (purple), where\begin{equation}
    v_\perp = \sqrt{v_R^2 + v_z^2}.
\end{equation} The thick disk/Splash stars preferentially hug the boundary of the velocity cut in the bottom right of each panel, where $V_{LSR}$ is the smallest. However, there are some thick disk/Splash stars that appear kinematically similar to the halo, and have large $v_\perp$ and small or negative $v_\phi$. Note that there are halo stars located at the same $v_\perp$ and $v_\phi$ as the thick disk/Splash stars. If one uses a more restrictive velocity cut instead of a chemical abundance cut to isolate halo stars, there will still be thick disk/Splash star contamination at low and negative $v_\phi$, and the data will lose a number of halo stars with positive $v_\phi$. 

The bottom row of Figure \ref{fig:disk_splash} shows $v_\phi$ and [Fe/H] for halo stars and thick disk/Splash stars, as well as the original boundary drawn over Splash stars by \cite{Belokurov2020}. Surprisingly, we find that the thick disk/Splash stars that we select with our chemical abundance cuts are shifted to lower [Fe/H] than was suggested by the \cite{Belokurov2020} selection box. \cite{Horta2021} also found Splash stars with [Fe/H] $<$ -0.8, and \cite{Donlon2022} claimed that there may be thick disk/Splash dwarf stars with [Fe/H] values as low as $-1$.

\cite{Amarante2020} claimed that a population of Splash stars can be generated by disk star+gas dynamics alone, without the need for a major merger event. However, that work was not able to exactly match the observed Splash star data; \cite{Belokurov2020} placed the lower $v_\phi$ boundary for Splash stars at -150 km s\invnospace, but the \cite{Amarante2020} model struggled to create Splash stars with $v_\phi < -50$ km s\invnospace. In our data, the observed thick disk/Splash star distribution appears to terminate near $v_\phi = -50$ km s\invnospace, which matches the cutoff of the \cite{Amarante2020} model, and is therefore consistent with a scenario in which the Splash star population was not created by a merger event.

Additionally, we do not see a bend in the $v_\phi$-[Fe/H] distribution of thick disk/Splash stars above $v_\phi > -100$ km s\invnospace; this is because our kinematic cuts removed the majority of disk stars, which dramatically outnumber Splash stars. Because \cite{Belokurov2020} drew their Splash star region over a row-normalized histogram, they were not able to identify Splash stars in the rows that contained disk stars. In our data, after the majority of disk stars have been removed, the thick disk/Splash star population actually extends up to $v_\phi > 200$ km s\invnospace.

\section{Chemodynamic Trends in Halo Stars} \label{sec:trends}


In this section we explore the expected chemodynamic distributions that would be generated from a single massive, ancient merger event. We then compare this model with observed data, and show that the energies and abundances of local halo stars are not consistent with this scenario. These observed chemodynamic trends in local halo stars can instead be explained by the local halo containing debris from multiple merger events. 

\subsection{Predictions for Chemodynamic Trends Due to a Single Massive, Ancient Merger Event}

\begin{figure*}
    \centering
    \includegraphics[width=\linewidth]{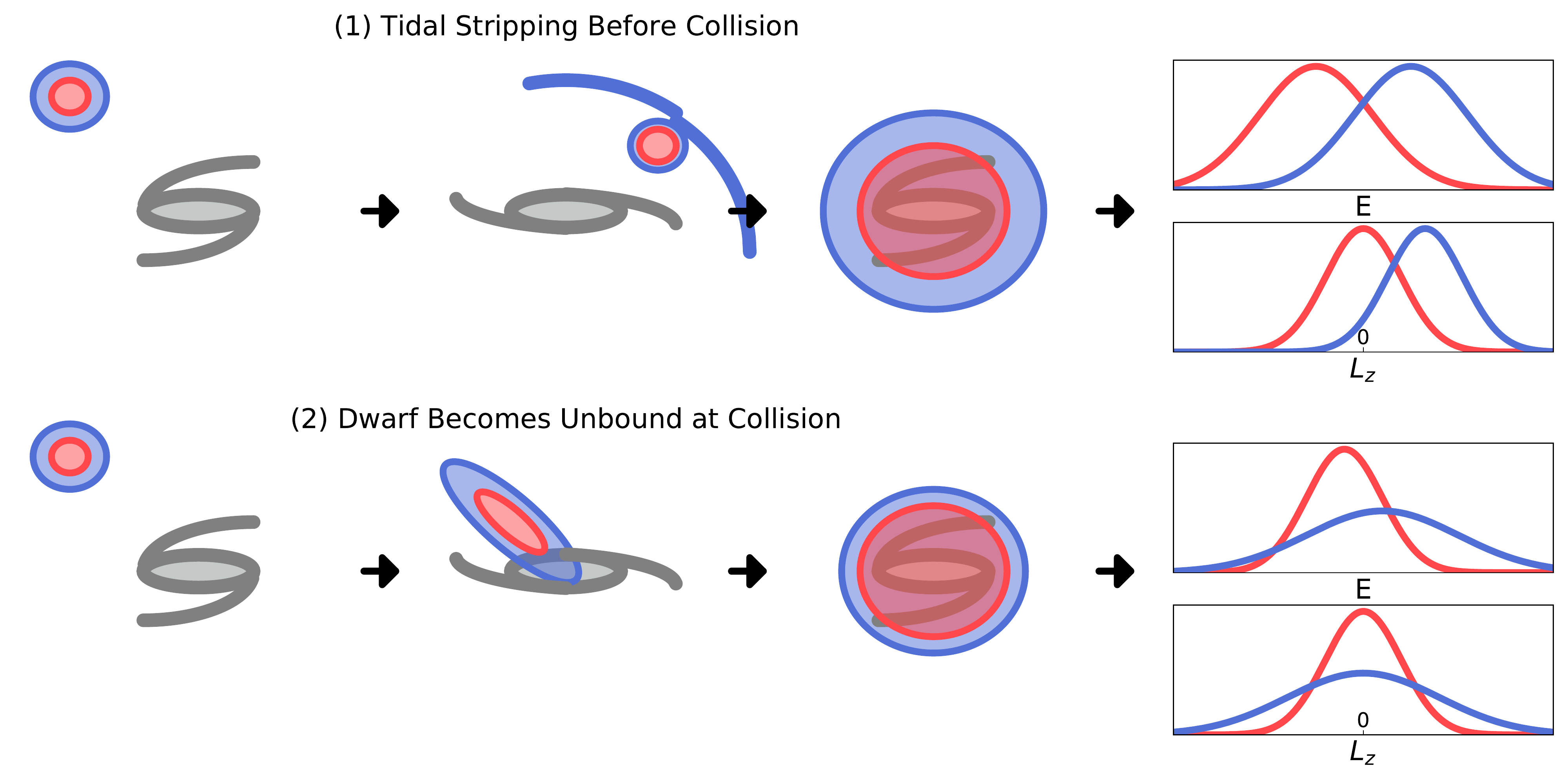}
    \caption{Two scenarios for an infall of a massive, ancient merger. The left part of the figure shows a drawing of the accretion in configuration space, increasing in time from left to right. The progenitor has an internal metallicity gradient; the blue material has low [Fe/H] and high \fe{\al}, and the red material has high [Fe/H] and low \fe{\al}. In the first scenario (top row), the outer portions of the progenitor are stripped to form an exaggerated tidal stream, while the bound core of the progenitor continues to spiral towards the host galaxy due to dynamical friction. In the second scenario (bottom row), the outer portions of the progenitor remain bound until collision with the host galaxy. At the end of both scenarios, there is a radial abundance gradient in the host galaxy halo stars that were deposited by the progenitor, but these stars have distinct chemodynamic trends shown in the right-most column. In reality, the infall of a dwarf galaxy will be somewhere between these two scenarios.}
    \label{fig:model}
\end{figure*}

Based on the current literature surrounding the GSE, we can surmise two generally accepted ideas about its progenitor dwarf galaxy: (i) it was massive, and (ii) it was accreted early on in the MW's history \citep{Belokurov2018,Helmi2018,Gallart2019,Kruijssen2020,Bonaca2020}. 

A massive progenitor informs us that the progenitor probably sustained enough star formation to generate a negative radial abundance gradient in its member stars. For example, Mercado et al. (2021) showed that in FIRE simulations, dwarf galaxies that assembled their stars early enough to have created the GSE have strong negative radial metallicity gradients. While \cite{MaiolinoMannucci2019} noted that there have been observations of positive metallicity gradients in galaxies, these are only weakly positive and are probably transient phenomena due to recent star formation, and therefore less likely to be good representatives of the ancient GSE merger. A massive progenitor (compared to the Milky Way at the time of infall) will also lose orbital energy due to dynamical friction.

The second point tells us that regardless of the progenitor's initial state, we should expect its debris to be well-mixed at the present day. With these assumptions in mind, we can build a qualitative model for the chemodynamic distributions of a single massive, ancient merger's member stars at the present day. 

Figure \ref{fig:model} shows two different scenarios for the accretion of a massive, ancient merger onto a MW-like galaxy. The two scenarios differ based on whether the outer regions of the progenitor become unbound before or after the progenitor dwarf galaxy becomes sufficiently radialized and collides with the host galaxy. 

In the first scenario, the outer portion of the progenitor with small [Fe/H] and large \fe{\al} is stripped before the bound core of the progenitor collides with the host galaxy. Once the outer portions are stripped from the progenitor, they can no longer lose energy and angular momentum via dynamical friction, so these quantities are locked in for the low-[Fe/H] stars. The bound progenitor core will continue falling in until it collides with the host galaxy, and will progressively lose more energy and angular momentum. At the present day, we expect to see a radial abundance gradient in the dwarf member stars, which are now located in the host galaxy. The low-[Fe/H] stars from the progenitor will have nonzero angular momentum and higher energy than the stars from the progenitor with large [Fe/H], which will have low energy and little-to-no angular momentum.

In the second scenario, the outer portion of the progenitor remains bound until the progenitor collides with the host galaxy. In this case, both the low and high [Fe/H] stars in the progenitor have the same angular momentum and energy on average. However, the low-[Fe/H] portion of the progenitor has a larger internal velocity dispersion and larger range of positions than the high-[Fe/H] portion of the progenitor. As a result, after the collision, the low-[Fe/H] stars will have a larger range of energy and angular momentum values with respect to the host galaxy. Additionally, the larger velocity dispersion of the low-[Fe/H] stars gives them somewhat higher energies after accretion.

In reality, the infall of a massive dwarf galaxy will be described by some mixture of these two scenarios. Some material may be stripped from the progenitor early on, and the progenitor may still have an abundance gradient at the time of impact. 

If the majority of the stars in the local stellar halo were deposited by a single massive, ancient merger event, those stars should follow the chemodynamic trends shown in Figure \ref{fig:model}. However, this is not the case; in the following subsections, we show that the observed chemodynamic trends in the local stellar halo are not consistent with either of these scenarios, or a mixture of scenarios.

\subsection{Energy \& Abundance Trends} \label{sec:e_at}

\begin{figure*}
    \centering
    \includegraphics[width=\linewidth]{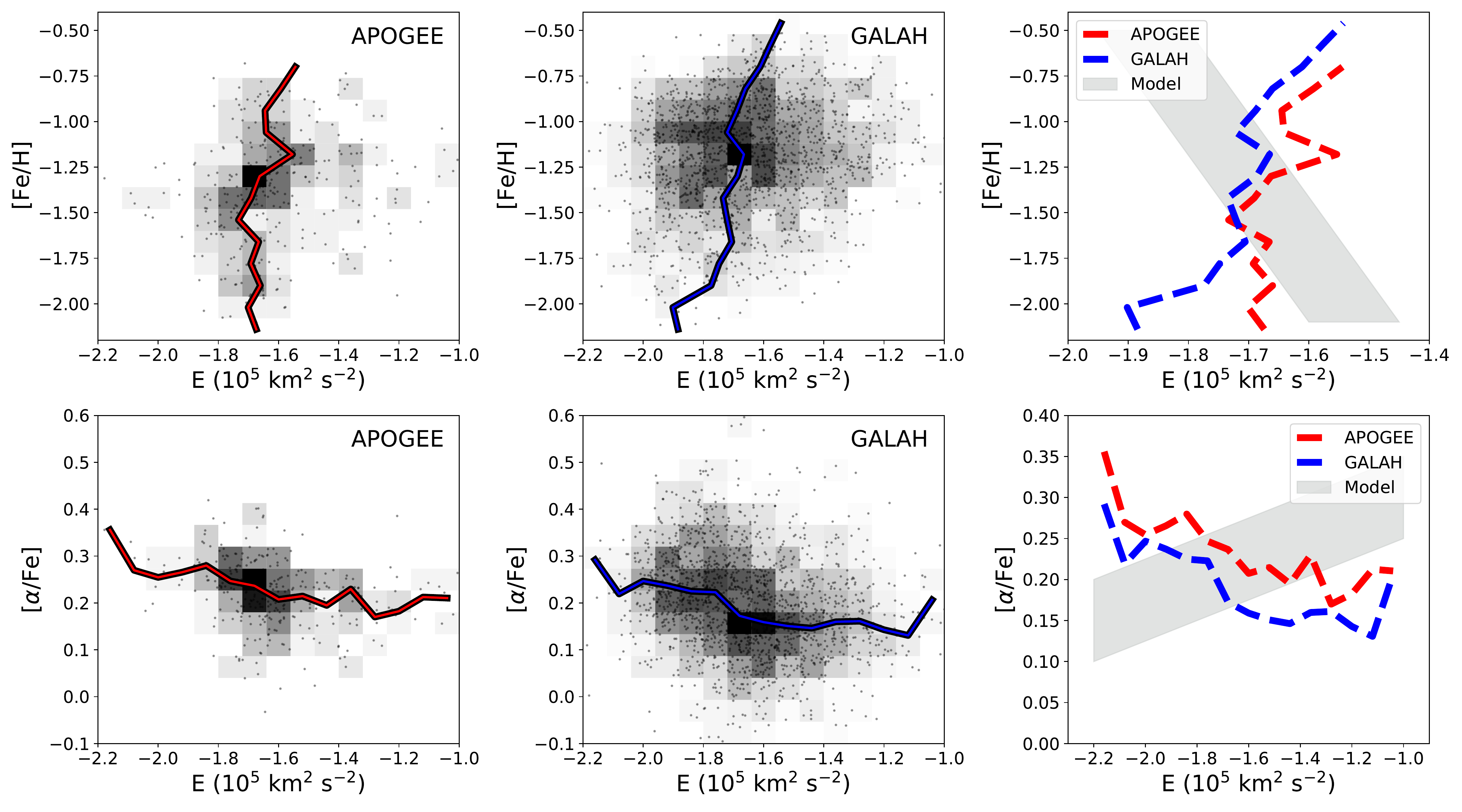}
    \caption{Heatmaps of abundances against orbital energy for halo stars. The top row shows [Fe/H] vs. energy, and the bottom row shows \fe{\al} vs. energy. The left column contains APOGEE data, the center column the GALAH data, and the right column shows the sigma-clipped averages of the observed data compared to the expected distribution from the models of the infall of a single massive dwarf galaxy at early times. In both datasets, there is an apparent trend where halo stars with larger energy have higher [Fe/H] values and lower \fe{\al} abundances. The observed chemodynamic trends are not consistent with our model for a single massive merger event such as the GSE. Rather, this trend can be interpreted as evidence of multiple radial merger events in the local stellar halo. Merger events that happen at early times would have lower [Fe/H] and high \fe{\al} abundances, and would be located deeper in the MW's gravitational potential well, which would give them lower energies. Merger events that happen at late times would have high [Fe/H] and low \fe{\al} abundances, and their stars would have higher energies than the stars from earlier mergers. }
    \label{fig:chem_e_trends}
\end{figure*}

\begin{figure*}
    \centering
    \includegraphics[width=\linewidth]{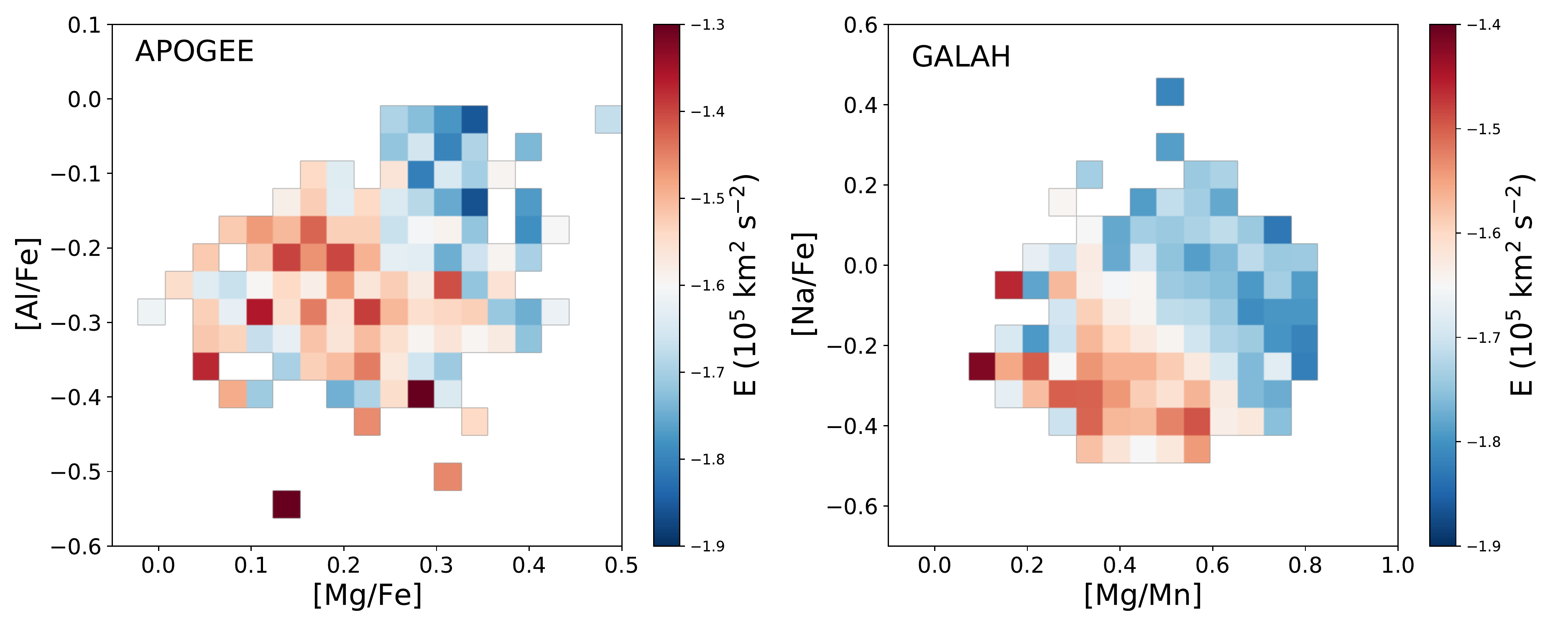}
    \caption{Heatmaps of orbital energy as a function of various chemical abundances for accreted stars with high eccentricity, which are primarily GSE stars. The top panel shows APOGEE stars plotted in \fe{Al} vs. \fe{Mg}, and the bottom panel shows GALAH stars plotted in \fe{Na} vs. [Mg/Mn]. Redder (bluer) bins show less (more) bound energies. The average energies of accreted halo stars appears to be correlated with their chemical abundances. The clustering of stars in this diagram indicates two things: (i) Overall, stars with different energies have different chemical abundances, which suggests that they formed in different environments, and (ii) Populations of stars with different energies can be separated with chemical cuts, and vice versa.}
    \label{fig:chem_energy}
\end{figure*}

Figure \ref{fig:chem_e_trends} shows the distributions of [Fe/H] and \fe{\al} abundances as functions of each star's orbital energy for the accreted APOGEE and GALAH stars that have orbital eccentricity $>$ 0.75. This is a reasonable criterion for selecting GSE member stars \citep[e.g.][]{Naidu2020}; thus, this figure only contains stars which are likely to belong to the GSE. As one looks at stars with progressively higher energies, the average [Fe/H] of those stars increases. Similarly, the stars with higher energy have lower average \fe{\al} values. 

In the right-most column of Figure \ref{fig:chem_e_trends}, we show the predicted energy-[Fe/H] and energy-\fe{\al} trends from Figure \ref{fig:model} in gray. On top of this expected trend, we show the sigma-clipped mean of the APOGEE and GALAH data as red and blue dashed lines, respectively. The observed data does not follow the predicted trend for [Fe/H] or \fe{\al}. While our model predicts a trend with a negative slope in [Fe/H] vs. energy, both the APOGEE and GALAH data show trends with positive slopes; similarly, our model predicts a positively-sloped trend in \fe{\al} vs. energy, but the APOGEE and GALAH data show negatively-sloping trends.

Similar energy-abundance trends are also present in other abundances than [Fe/H] and \fe{\al}. Figure \ref{fig:chem_energy} shows heatmaps of orbital energy as a function of various chemical abundances in the APOGEE and GALAH data. Overall, stars with larger values of \fe{X} abundances have lower energy than stars with low \fe{X} abundances, which is the opposite of the expected trend based on our single merger model. Other works have previously found a connection between the chemical abundances of halo stars and their dynamics; our results are similar to those of \cite{Das2020} and \cite{Ness2022}, who also used \fe{Mg} and \fe{Al} in order to isolate distinct halo structures in APOGEE data, and \cite{Buder2022}, who used \fe{Na} and [Mg/Mn] in their analysis of GALAH stars.

We conclude that the energy/abundance trends in the observed data are inconsistent with the predicted chemodynamic trends of the GSE; the observed data can only be explained if the progenitor galaxy had a strong positive metallicity gradient, which is unlikely for an early, massive merger. However, it is possible to explain the observed chemodynamic trends if the local solar region contains debris from several distinct radial merger events. Dwarf galaxies have a range of chemical abundances, and when they are accreted, the energy of their constituent material will depend on many factors such as the mass, orbital parameters, and infall time of the progenitor dwarf galaxy. If, for example, a low-[Fe/H], high-\fe{\al} dwarf galaxy fell in at early times (so that its material would have low-energy at the present day), and a high-[Fe/H], low-\fe{\al} dwarf galaxy fell in recently (so that its material would have high energy), then that could explain the observed energy-abundance trends. Notably, this scenario is consistent with the MW accretion history proposed by \cite{Donlon2022}. 

\subsection{$L_z$-Abundance Trends}

In principle, it should be possible to also evaluate whether the $L_z$-abundance trends of the observed data are consistent with a single massive, ancient merger scenario. According to our model, one should see that the local halo stars with $L_z\sim0$ should have higher [Fe/H] and smaller \fe{\al} than the local halo stars with nonzero $L_z$. Unfortunately, this is difficult in practice, because different reasonable choices of the Sun's kinematic frame can shift observed $L_z$ values by upwards of 200 kpc km s\inv (see Appendix \ref{app:lz_diff}). Because stars in the radial halo have $L_z$ values on the order of hundreds of kpc km s\invnospace, this means that stars with zero $L_z$ in one choice of frame can have significantly nonzero $L_z$ in another frame, and vice versa; this makes it nontrivial to determine what stars have $L_z\sim0$ and which have nonzero $L_z$. 

Despite this limitation, it is still possible to compare relative $L_z$-abundance information with our model. \cite{Donlon2022} showed that the low-[Fe/H] dwarf stars in the local stellar halo have more prograde velocities than the nearby high-[Fe/H] dwarf stars. If one assumes that the high-[Fe/H] stars have $L_z=0$ in order to be consistent with our single massive, ancient merger model, then the low-[Fe/H] stars would have prograde orbits. If these two populations were generated from the same infall event, then this implies the first scenario in Figure \ref{fig:model}, because there is not a corresponding collection of low-[Fe/H] stars with retrograde orbits. This would require that the infall of the progenitor was on a prograde orbit, which is not consistent with the findings of \cite{Naidu2021} that the progenitor of the GSE must have fallen in on a retrograde orbit.

In this way, the $L_z$-[Fe/H] distribution of the local stellar halo stars is also inconsistent with the current understanding of the GSE. However, the $L_z$-[Fe/H] distribution of nearby halo stars can be explained if there were multiple distinct merger events with different $L_z$ and [Fe/H] distributions that were accreted by the MW.

\section{Gaussian Mixture Model Classification}  \label{sec:model_fit}

The chemodynamic trends observed in Section \ref{sec:trends} suggest that it should be possible to separate different halo structures in the solar neighborhood using multi-dimensional combinations of dynamical quantities and chemical abundances. In this section, we investigate this idea further, using the Gaussian Mixture Model (GMM) implementation from the \verb!scikit-learn! python package \citep{scikit-learn} to split the data into multidimensional Gaussian components. We then analyze the identified GMM components, and match them to known halo substructure based on their chemodynamic properties. 

GMMs can struggle to correctly identify the correct number and shapes of components in a sample if they are fitting quantities that are correlated with each other. This is because if two quantities are individually normally distributed, but share some correlation, then their joint distribution is no longer expected to be a multidimensional normal distribution. Notably, this is true for orbital energy $E$ and $L_z$, which are both functions of $v_\phi$. In order to remove the correlation between the quantities in our fitting method, we chose to replace orbital energy with a ``pseudo-energy'' $\widetilde{E}$, defined as \begin{equation}
\widetilde{E} = \frac{\mathbf{v}^2 - v^2_\phi}{2} + \Phi(\mathbf{x}).
\end{equation} Provided that $v_\phi$ is small compared to the total velocity of each star, this quantity will be roughly an integral of the motion.

We took the quantities $\widetilde{E}$, $L_z$, [Fe/H], \fe{\al}, \fe{Na}, and \fe{Al} for each star in the APOGEE sample, and then optimized a single component GMM over this data. We then calculated the Bayesian Information Criteria \citep[BIC, ][]{bic} for this fit using the built-in \verb!scikit-learn! BIC function for GMMs.

\begin{figure*}
    \centering
    \includegraphics[width=\linewidth]{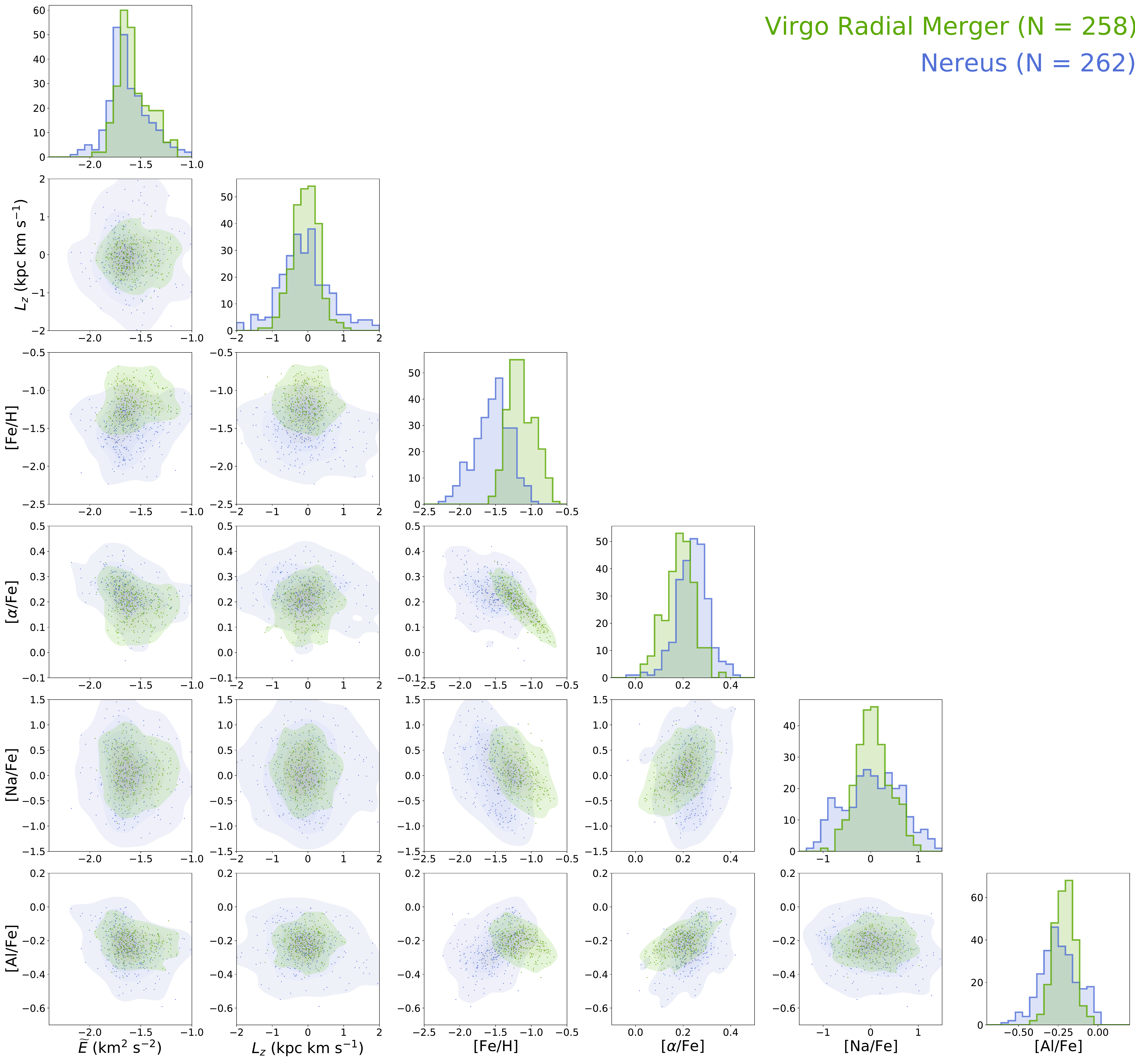}
    \caption{The best-fit GMM decomposition of kinematic and chemical properties of APOGEE stars. This fit contains two components, denoted by the colors blue and green. Based on dynamical and chemical properties, the green component corresponds to the VRM, and the blue component corresponds to Nereus. All stars in the data were assigned to one of these components, and are colored accordingly. Each row and column correspond to one of the six quantities that were used in the fitting procedure: the top panel of each column shows a histogram of that quantity split up by component, and each other panel shows a scatter plot plus shaded contours, colored by component.}
    \label{fig:gmm_apogee}
\end{figure*}

Once we had the BIC for the single component GMM, then we fit a two component GMM on the same data and recalculated the BIC for the two component fit. The BIC is a heuristic that is designed to evaluate the trade-off between the improvement in the likelihood score of a model fit with more parameters and the number of parameters that were added to that new model. A decrease in the BIC indicates a better model. We repeated this process for up to seven GMM components, and then compared the BIC of each model. The BIC was minimized by the fit with two components, so we adopt the two component model as our best fit. The results of this fit are shown in Figure \ref{fig:gmm_apogee}. 

\begin{figure*}
    \centering
    \includegraphics[width=\linewidth]{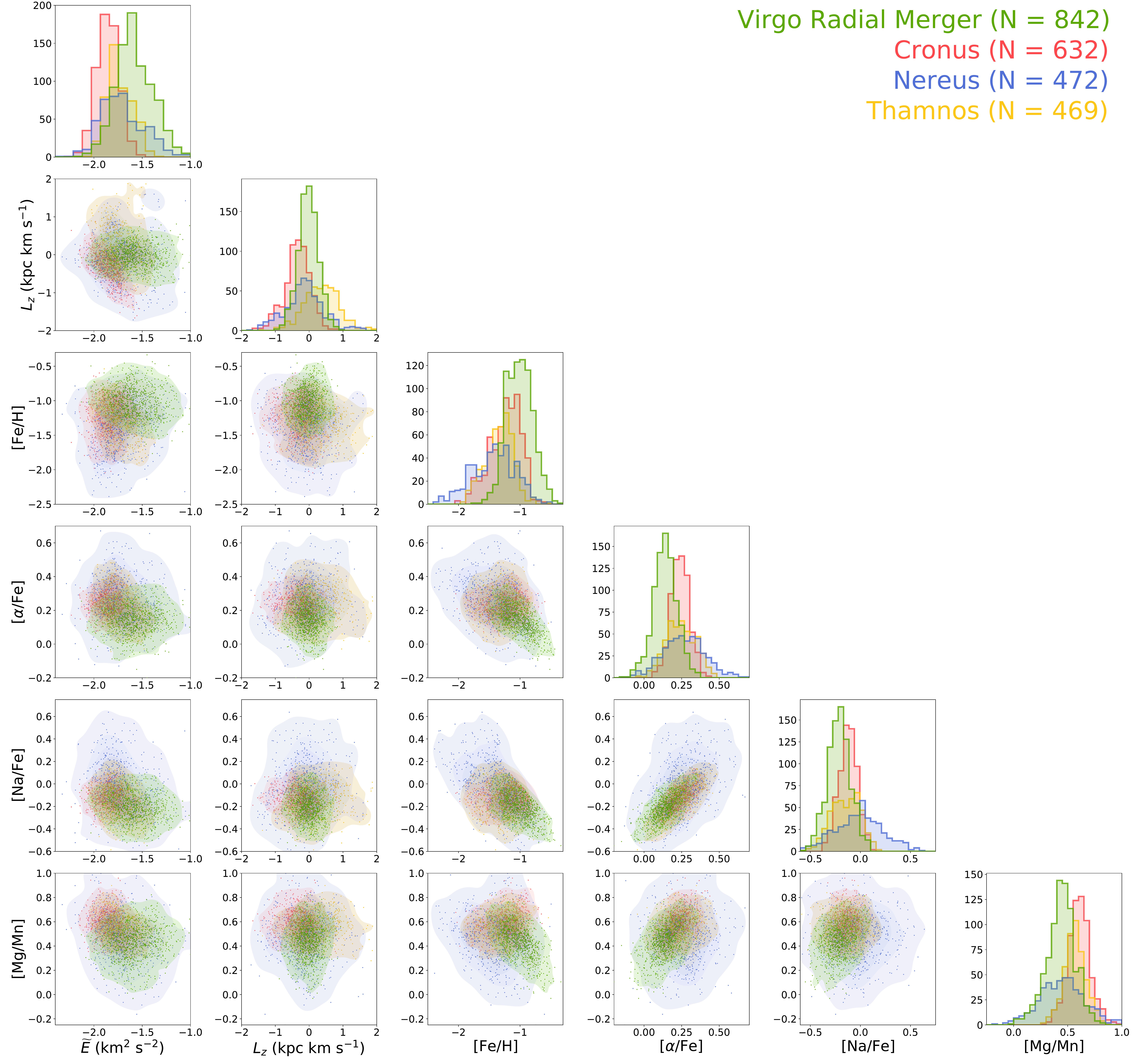}
    \caption{The four component GMM decomposition of kinematic and chemical properties of GALAH stars. The different components are denoted by the colors red, blue, green, and yellow. Based on kinematic and chemical properties, the green component corresponds to the VRM, the red component corresponds to Cronus, the blue component corresponds to Nereus, and the yellow component corresponds to Thamnos. The plotted stars are colored according to which halo component the GMM regression algorithm determined was the best fit for that star. Each row and column correspond to one of the six quantities that were used in the fitting procedure: the top panel of each column shows a histogram of that quantity split up by component, and each other panel shows a scatter plot plus shaded contours, colored by component.}
    \label{fig:gmm_galah}
\end{figure*}

We then repeated this procedure for the GALAH sample, except that \fe{Al} was replaced by [Mg/Mn], because fewer GALAH stars have high-quality \fe{\al} measurements. The GMM model with four components had the lowest BIC for the GALAH data. The results of the four-component GMM fit for the GALAH data are shown in Figure \ref{fig:gmm_galah}. 

Each star in the APOGEE and GALAH samples was then assigned to one of the fit components in the corresponding GMM fit. The probability $p_k$ that a star with parameters $\mathbf{s}$ belongs to the $k$th Gaussian component is defined as \begin{equation}
p_k = \frac{f_k(\mathbf{s})}{\sum_i^4 f_i(\mathbf{s})},
\end{equation} where $f_k(\mathbf{s})$ is the evaluation of the multivariate normal distribution of the $k$th component at $\mathbf{s}$. 

This will separate the stars into the components from which they are actually associated only if the components are not overlapping in the six dimensions that we use to separate them. If the components are overlapping, then we only recover the underlying distributions, not the actual associations for each star. Here, we only need the distribution of each component in our six chemodynamical parameters to identify the properties of each component and the structure(s) to which each component belongs.

To explore this idea further, we assigned a ``confidence'' value to each star, defined as the maximum $p_k$ that a given star had for all components. This confidence value can be interpreted as the likelihood that each given star was assigned to the correct GMM component, assuming that the GMM components are known exactly. For the APOGEE data the mean confidence value is 90.3\%, and the median confidence value is 96.9\%. For the GALAH data, the mean confidence value is 79.3\%, and the median confidence value is 83.3\%. We tried removing all stars with confidence values below an arbitrary threshold value (75\% and 90\%) from the data when performing our analysis, but the overall high confidence values for the stars in the model meant that removing these stars didn't substantially change any results.

Appendix \ref{app:mock_fit} provides an assessment of the algorithm's ability to identify individual dwarf galaxies from a mixed distribution, based on a fit of known MW dwarf galaxy data. The algorithm is able to accurately separate the mixed data into the constituent dwarf galaxies, so we are confident that the algorithm is able to identify distinct dwarf galaxy components in the nearby stellar halo data as well.

\subsection{Identifying the Separated Components} \label{sec:naming}

Three distinct components of the local halo dwarf stars were identified by \cite{Donlon2022}, which are named the VRM, Nereus, and Cronus. \cite{Donlon2022} determined kinematic and metallicity distributions for each structure: they found that the VRM has the largest energy of the structures, followed by Nereus, which has a slightly lower energy than the VRM, and then Cronus, which had a much lower energy than the other two structures. The rotational velocities of these structures were also different: The VRM was somewhat retrograde, Nereus was non-rotating, and Cronus had prograde velocity. \cite{Donlon2022} also show that these structures have different photometrically-determined metallicities: The VRM has [Fe/H]$_{\textrm{phot}}$ $\sim$ -1.7, Nereus has [Fe/H]$_{\textrm{phot}}$ $\sim$ -2.1, and Cronus has [Fe/H]$_{\textrm{phot}}$ $\sim$ -1.2. 

In our GMM fits, we identify two components of the local stellar halo in the APOGEE data, and four components of the local stellar halo in the GALAH data. The GALAH data contains the two components from the APOGEE data plus two additional components not present in the APOGEE data; the differences between these fits are discussed in Section \ref{sec:vs_fit}. The first halo component, shown in green, contains stars with high energy and small $L_z$ values. This component has a high [Fe/H] value, but has low values of \fe{\al}. These properties are consistent with the VRM, which has been shown to have large energy and a relatively high [Fe/H] compared to the surrounding stellar halo \citep{Donlon2019,Donlon2022}. 

The \fe{\al} abundance distribution of stars within a halo component contains information about the star formation history of the population. This is because Type-II supernovae (SNe II) happen early-on in star formation histories, and enrich the surrounding gas with large amounts of \als elements compared to Fe. This results in large \fe{\al} for the stars formed out of the material from SNe II. Type-Ia supernovae (SNe Ia) become more common after multiple epochs of star formation, and enrich the surrounding gas with large amounts of Fe and Mn, which depletes the \fe{\al} abundance of stars that form out of SNe Ia material. Based on this, the low \fe{\al} abundances of the green stars in the GMM fits indicate that their progenitor dwarf galaxy sustained active star formation for a long period of time. A long period of star formation is consistent with a dwarf galaxy only being accreted and subsequently disrupted recently in the Galaxy's history, which \cite{Donlon2019} and \cite{Donlon2020} argue is the case for the VRM. We therefore conclude that the green component corresponds to the VRM.

The second component, shown in red in Figure \ref{fig:gmm_galah}, contains stars with low energy, and prograde $L_z$. Additionally, this component has a high [Fe/H] abundance, although it is not quite as [Fe/H]-rich as the VRM. The energy and $L_z$ of this component corresponds to the properties of the Cronus merger event as described in \cite{Donlon2022}, so we argue that the red component corresponds to the Cronus merger event. 

We find that Cronus has high \fe{\al}, and slightly higher \fe{Na} and [Mg/Mn] abundances than the VRM. Cronus' high \fe{\al} and [Mg/Mn] abundances indicate that its star formation ended before rates of SNe Ia had time to overtake SNe II rates. This is consistent with an early time of accretion into the MW, which is in agreement with the analysis of \cite{Donlon2022}, who pointed to Cronus' low energy as evidence of an early time of accretion.

The third component, shown in blue, contains stars with energies smaller than those of the VRM stars, but larger than the energies of the Cronus stars. The stars in this component have essentially no overall rotation. This component has low [Fe/H] abundances, high \fe{Mg} abundances, and high \fe{Na} abundances. The relatively large energy, lack of overall rotation, and low [Fe/H] abundance of this component leads us to conclude that this component corresponds to the Nereus structure from \cite{Donlon2022}. 

In general, the chemical abundance distributions for Nereus stars have a larger dispersion than the chemical abundance distributions for the VRM stars or Cronus stars. This may indicate that the collection of stars we call Nereus is not from a single merger event, but instead is formed from many minor radial mergers that were independently accreted onto the MW. One might expect that each minor merger would have a slightly different distribution of chemical abundances with a small dispersion; if all of these minor mergers were attributed to a single component of the stellar halo, that component would have a large apparent dispersion in its chemical abundance distribution, even though each individual minor merger has a small dispersion in chemical abundances. In contrast, the shapes of the chemical abundance distributions for the VRM and Cronus components appear similar to those seen in the classical dwarf galaxies \citep{Kirby2013}. We revisit the idea that Nereus may be formed out of multiple components in Section \ref{sec:nature_nereus}.

This leaves a final component, in yellow, which has stars with relatively low energy and retrograde $L_z$. These stars have relatively high [Fe/H] and \fe{\al}, as well as fairly high [Mg/Mn] and \fe{Na} abundances. This component is consistent with the Thamnos structure \citep{Koppelman2019b}, which was originally discovered in APOGEE, RAVE, and LAMOST stars. This original discovery restricted their sample to within 3 kpc of the Sun, so Thamnos is expected to be present in the local stellar halo. Interestingly, \cite{Koppelman2019b} identify two distinct halves of the Thamnos structure, which they claim are from the same progenitor dwarf galaxy. We are able to identify different chemical structures in Thamnos as well (Section \ref{sec:nature_nereus}), which may indicate that these distinct halves of Thamnos are actually the result of material from multiple progenitors.

Notably, while the VRM component does contain the most stars, the VRM only contains roughly one third of all of the stars in the GALAH dataset. This means that the VRM does not dominate the local stellar halo, as the GSE is claimed to do. Rather, each of the identified components make a sizeable contribution to the local stellar halo. 

\subsection{Differences Between the APOGEE \& GALAH Fits} \label{sec:vs_fit}

The best GMM fit for the APOGEE data contains two components (the VRM and Nereus), while the best GMM fit for the GALAH data contains four components (the VRM and Nereus, plus Cronus and Thamnos). The primary reason for the different number of components in the two samples is due to the spatial volumes of the two surveys; because the APOGEE data is mainly located in the directions of the Galactic caps, while the GALAH data is mainly located in the direction of the Galactic center, this causes a difference in the energy ranges probed by the two surveys \citep{Lane2022}.

Our APOGEE data only extends down to $\widetilde{E}\sim-1.8\times10^5$ km$^2$ s$^{-2}$, while the GALAH data extends down to $\widetilde{E}\sim-2.0\times10^5$ km$^2$ s$^{-2}$. The mean energies of Cronus and Thamnos are below $\widetilde{E}\sim-1.8\times10^5$ km$^2$ s$^{-2}$, but the mean energies of the VRM and Nereus are above this threshold. This generates a selection effect where the majority of stars in the APOGEE data belong to the VRM and Nereus, which in turn causes the GMM algorithm to only identify the VRM and Nereus components in the APOGEE data. This selection effect is not present in the GALAH data, so the GMM algorithm is able to identify two additional components in the GALAH data (Cronus and Thamnos) as well as the two components identified in the APOGEE data.

In addition, the properties of the VRM and Nereus are slightly different in the APOGEE data vs the GALAH data (see Section \ref{sec:characteristics}). This is probably due to the presence of a small number of Cronus and Thamnos stars in the APOGEE data, which are mixed into the identified VRM and Nereus components, but there are not enough stars to allow the GMM algoritm to identify the Cronus and Thamnos structures. Alternatively, these differences may be due to variance in the properties of each substructure over different volumes, which could possibly indicate that the local stellar halo substructures are not in dynamical equilibrium, although disequilibrium is not required to generate spatial chemodynamic trends from accretion events (i.e. Figure \ref{fig:model}).

\section{Characteristics of Each Merger Event} \label{sec:characteristics}

Our GMM fits only utilize two kinematic and four chemical quantities for each star. However, our data contains many more dynamical quantities and chemical abundances than those used in the fitting procedure. In this section, we look at additional properties of the four components of the stellar halo. 

The means and standard deviations of each dynamical quantity and chemical abundance for the identified halo structures are given in Table \ref{tab:properties}.

\begin{deluxetable*}{lrrrrrr}
\tablecaption{Means and standard deviations of kinematic, dynamical, and chemical properties of stars assigned to each merger event, separated by spectroscopic survey. 
    \label{tab:properties}}
\tablehead{ & \multicolumn{2}{c}{\textbf{Virgo Radial Merger}} & \textbf{Cronus} & \multicolumn{2}{c}{\textbf{Nereus}} & \textbf{Thamnos} \\
           Quantity & APOGEE & GALAH & GALAH & APOGEE & GALAH & GALAH} 
\startdata
\textbf{$|v_R|$} (km/s)                  &      156$\pm$89 &     199$\pm$94 &      91$\pm$61 &      136$\pm$95 &     147$\pm$97 &     126$\pm$80 \\
\textbf{$v_\phi$} (km/s)                 &        7$\pm$43 &       4$\pm$44 &      54$\pm$63 &        9$\pm$91 &      24$\pm$90 &     -55$\pm$80 \\
$v_z$ (km/s)                             &        3$\pm$92 &      -8$\pm$85 &       7$\pm$86 &        1$\pm$93 &      -2$\pm$84 &       5$\pm$76 \\
Eccentricity                             &   0.73$\pm$0.20 &  0.91$\pm$0.09 &  0.74$\pm$0.18 &   0.69$\pm$0.23 &  0.79$\pm$0.18 &  0.74$\pm$0.20 \\
$z_\textrm{max}$ (kpc)                   &     7.0$\pm$4.8 &    5.3$\pm$5.7 &    3.1$\pm$1.7 &    7.5$\pm$13.4 &    4.4$\pm$4.6 &    3.6$\pm$2.5 \\
$r_m$ (kpc)                              &     7.6$\pm$2.8 &    7.4$\pm$3.7 &    4.5$\pm$1.5 &     8.3$\pm$7.6 &    6.8$\pm$4.4 &    5.9$\pm$2.1 \\
$r_{apo}$ (kpc)                          &    13.1$\pm$4.7 &   14.3$\pm$7.2 &    7.7$\pm$2.2 &   14.1$\pm$14.4 &   12.2$\pm$8.2 &   10.3$\pm$3.7 \\
$r_{peri}$ (kpc)                         &     2.2$\pm$1.9 &    0.6$\pm$0.7 &    1.3$\pm$1.2 &     2.6$\pm$2.3 &    1.4$\pm$1.4 &    1.5$\pm$1.3 \\
\textbf{Energy (10$^5$ km$^2$ $s^{-2}$)} &  -1.56$\pm$0.16 & -1.57$\pm$0.20 & -1.82$\pm$0.14 &  -1.57$\pm$0.23 & -1.66$\pm$0.24 & -1.69$\pm$0.14 \\
\textbf{$L_z$ (kpc km s$^{-1}$)}         &     -64$\pm$368 &    -24$\pm$298 &   -347$\pm$383 &     -64$\pm$744 &   -146$\pm$610 &    371$\pm$528 \\
$J_R$ (kpc km s$^{-1}$)                  &     631$\pm$368 &   1023$\pm$583 &    358$\pm$157 &     677$\pm$970 &    715$\pm$630 &    547$\pm$319 \\
$J_z$ (kpc km s$^{-1}$)                  &     391$\pm$440 &    174$\pm$219 &    174$\pm$152 &     350$\pm$416 &    174$\pm$209 &    153$\pm$170 \\
\textbf{\textrm{[Fe/H]}} & -1.14$\pm$0.19 & -1.01$\pm$0.21 & -1.24$\pm$0.25 & -1.54$\pm$0.25 & -1.43$\pm$0.36 & -1.34$\pm$0.23 \\
\textbf{\fe{\al}}        &  0.18$\pm$0.06 &  0.14$\pm$0.08 &  0.24$\pm$0.06 &  0.24$\pm$0.07 &  0.27$\pm$0.15 &  0.24$\pm$0.10 \\
\fe{Li}                  &              - &  1.40$\pm$0.71 &  1.37$\pm$0.48 &              - &  1.99$\pm$0.67 &  1.61$\pm$0.62 \\
\fe{C}                   & -0.34$\pm$0.16 &              - &              - & -0.28$\pm$0.27 &              - &              - \\
\fe{CI}                  & -0.20$\pm$0.20 &              - &              - & -0.10$\pm$0.26 &              - &              - \\
\fe{N}                   &  0.14$\pm$0.13 &              - &              - &  0.24$\pm$0.26 &              - &              - \\
\fe{O}                   &  0.31$\pm$0.12 &  0.50$\pm$0.22 &  0.62$\pm$0.22 &  0.37$\pm$0.12 &  0.67$\pm$0.32 &  0.64$\pm$0.20 \\
\textbf{\fe{Na}}         &  0.05$\pm$0.35 & -0.21$\pm$0.12 & -0.12$\pm$0.09 &  0.05$\pm$0.59 & -0.01$\pm$0.12 & -0.15$\pm$0.14 \\
\textbf{\fe{Mg}}         &  0.17$\pm$0.08 &  0.10$\pm$0.10 &  0.23$\pm$0.09 &  0.25$\pm$0.08 &  0.15$\pm$0.16 &  0.20$\pm$0.11 \\
\textbf{\fe{Al}}         & -0.21$\pm$0.06 & -0.06$\pm$0.21 &  0.13$\pm$0.16 & -0.24$\pm$0.12 &  0.32$\pm$0.38 &  0.20$\pm$0.24 \\
\fe{Si}                  &  0.19$\pm$0.06 &  0.12$\pm$0.11 &  0.24$\pm$0.11 &  0.23$\pm$0.07 &  0.28$\pm$0.19 &  0.25$\pm$0.14 \\
\fe{S}                   &  0.37$\pm$0.14 &              - &              - &  0.38$\pm$0.24 &              - &              - \\
\fe{K}                   &  0.18$\pm$0.16 &  0.09$\pm$0.15 &  0.17$\pm$0.16 &  0.21$\pm$0.32 &  0.14$\pm$0.20 &  0.14$\pm$0.14 \\
\fe{Ca}                  &  0.17$\pm$0.08 &  0.20$\pm$0.11 &  0.28$\pm$0.09 &  0.19$\pm$0.18 &  0.29$\pm$0.15 &  0.27$\pm$0.11 \\
\fe{Sc}                  &              - &  0.07$\pm$0.10 &  0.10$\pm$0.09 &              - &  0.10$\pm$0.13 &  0.10$\pm$0.10 \\
\fe{Ti}                  & -0.06$\pm$0.11 &  0.20$\pm$0.16 &  0.26$\pm$0.14 & -0.05$\pm$0.21 &  0.37$\pm$0.29 &  0.28$\pm$0.21 \\
\fe{TiII}                &  0.19$\pm$0.23 &  0.30$\pm$0.13 &  0.37$\pm$0.12 &  0.12$\pm$0.29 &  0.39$\pm$0.17 &  0.36$\pm$0.13 \\
\fe{V}                   & -0.03$\pm$0.37 &  0.05$\pm$0.33 & -0.01$\pm$0.26 &  0.10$\pm$0.42 &  0.17$\pm$0.56 &  0.00$\pm$0.36 \\
\fe{Cr}                  & -0.20$\pm$0.30 & -0.14$\pm$0.13 & -0.15$\pm$0.13 & -0.10$\pm$0.38 & -0.06$\pm$0.22 & -0.13$\pm$0.15 \\
\textbf{\fe{Mn}}         & -0.34$\pm$0.12 & -0.33$\pm$0.10 & -0.37$\pm$0.09 & -0.34$\pm$0.20 & -0.30$\pm$0.19 & -0.37$\pm$0.11 \\
\fe{Co}                  & -0.21$\pm$0.34 &  0.07$\pm$0.45 &  0.20$\pm$0.45 & -0.10$\pm$0.48 &  0.50$\pm$0.75 &  0.24$\pm$0.54 \\
\fe{Ni}                  & -0.05$\pm$0.12 & -0.16$\pm$0.12 & -0.12$\pm$0.13 & -0.06$\pm$0.11 & -0.09$\pm$0.25 & -0.13$\pm$0.15 \\
\fe{Cu}                  &              - & -0.49$\pm$0.14 & -0.40$\pm$0.16 &              - & -0.41$\pm$0.29 & -0.48$\pm$0.18 \\
\fe{Zn}                  &              - &  0.16$\pm$0.18 &  0.23$\pm$0.18 &              - &  0.23$\pm$0.19 &  0.22$\pm$0.18 \\
\fe{Y}                   &              - &  0.12$\pm$0.27 &  0.25$\pm$0.30 &              - &  0.12$\pm$0.33 &  0.18$\pm$0.30 \\
\fe{Zr}                  &              - &  0.44$\pm$0.47 &  0.55$\pm$0.38 &              - &  0.67$\pm$0.69 &  0.64$\pm$0.47 \\
\fe{Ba}                  &              - &  0.31$\pm$0.34 &  0.36$\pm$0.37 &              - &  0.25$\pm$0.42 &  0.31$\pm$0.36 \\
\fe{La}                  &              - &  0.40$\pm$0.33 &  0.39$\pm$0.31 &              - &  0.59$\pm$0.50 &  0.42$\pm$0.32 \\
\fe{Ce}                  & -0.19$\pm$0.28 & -0.05$\pm$0.32 & -0.10$\pm$0.32 & -0.23$\pm$0.39 &  0.13$\pm$0.46 & -0.04$\pm$0.35 \\
\fe{Nd}                  &              - &  0.56$\pm$0.22 &  0.48$\pm$0.23 &              - &  0.64$\pm$0.35 &  0.53$\pm$0.24 \\
\fe{Sm}                  &              - &  0.26$\pm$0.31 &  0.19$\pm$0.33 &              - &  0.43$\pm$0.49 &  0.22$\pm$0.32 \\
\fe{Eu}                  &              - &  0.49$\pm$0.15 &  0.40$\pm$0.20 &              - &  0.52$\pm$0.27 &  0.47$\pm$0.28
\enddata
\tablecomments{Bold font indicates that the quantity was used in the GMM fit to identify the components.}
\end{deluxetable*}

\subsection{Dynamics}

\begin{figure*}
    \centering
    \includegraphics[width=\linewidth]{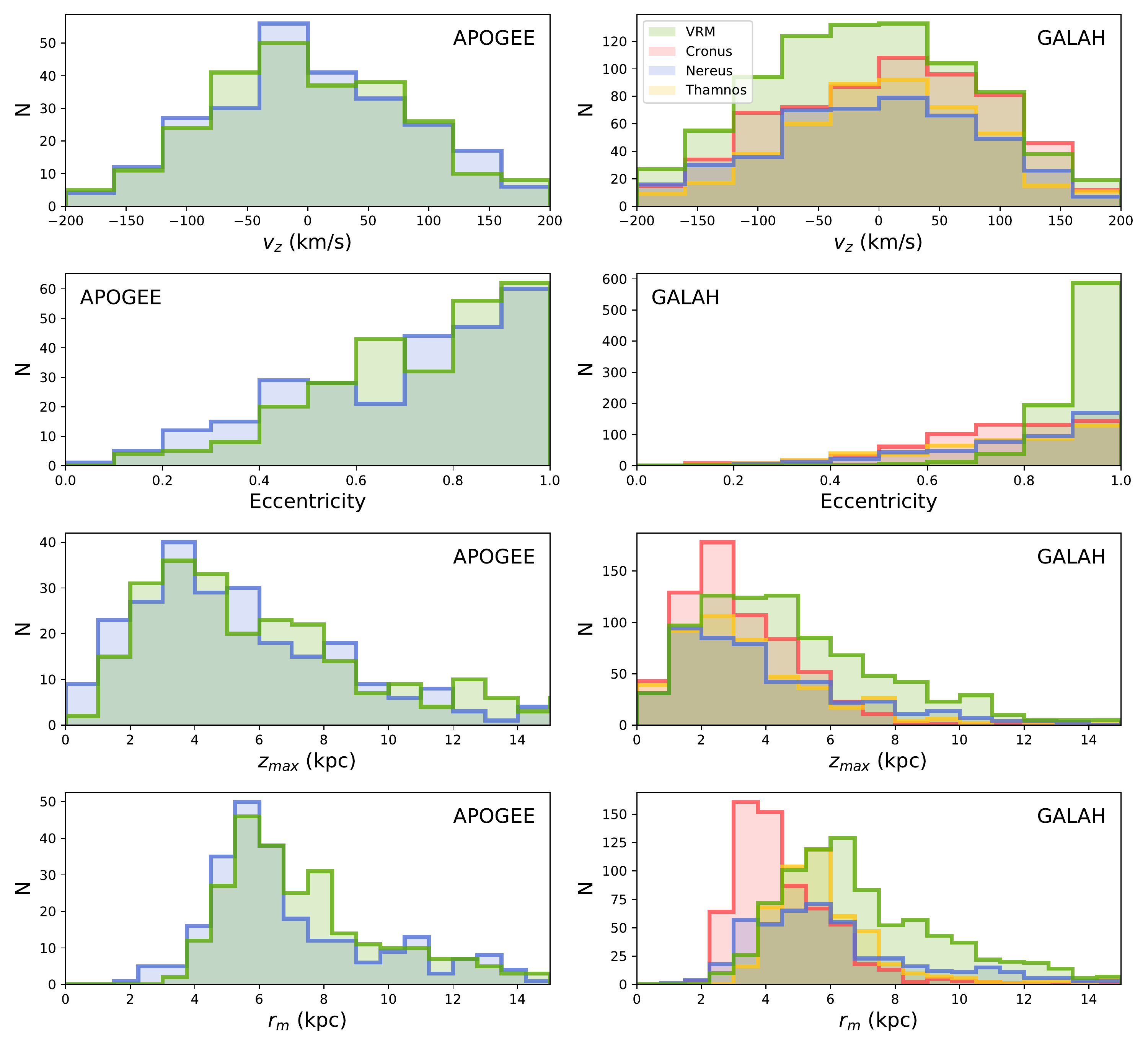}
    \caption{Dynamical properties of the GMM components. Green corresponds to the VRM, red corresponds to Cronus, blue corresponds to Nereus, and yellow corresponds to Thamnos. The left column shows the APOGEE sample, and the right column shows the GALAH sample. The top row contains Galactocentric Cartesian $Z$ velocity, the second row contains orbital eccentricity, the third row contains maximum orbital distance from the Galactic plane ($z_\textrm{max}$), and the bottom row contains the mean orbital distance from the Galactic center ($r_m$). The $v_z$ distributions of each component change between the APOGEE sample and the GALAH sample; this may be due to systematic effects related to the different footprints of the two surveys, or it may indicate that the $v_z$ distributions of each component actually vary as a function of position. The majority of stars in both samples have eccentricities larger than 0.6. For APOGEE stars, the VRM may have a somewhat larger eccentricity than Nereus stars, although there is not a large difference between the two components. In the GALAH sample, the VRM has a much larger orbital eccentricity than the other components, which have comparable eccentricity distributions, although Nereus appears somewhat more radial than Cronus and Thamnos. The VRM and Nereus appear to have similar $z_\textrm{max}$ distributions in the APOGEE data; however, in the GALAH data, the VRM appears to have a larger $z_\textrm{max}$ values than the other components. Finally, in both samples Cronus appears to have the lowest $r_m$ values, the VRM stars have the highest $r_m$ values, and Nereus and Thamnos stars have values of $r_m$ somewhere in between.}
    \label{fig:dynamics}
\end{figure*}

\begin{figure*}
    \centering
    \includegraphics[width=\linewidth]{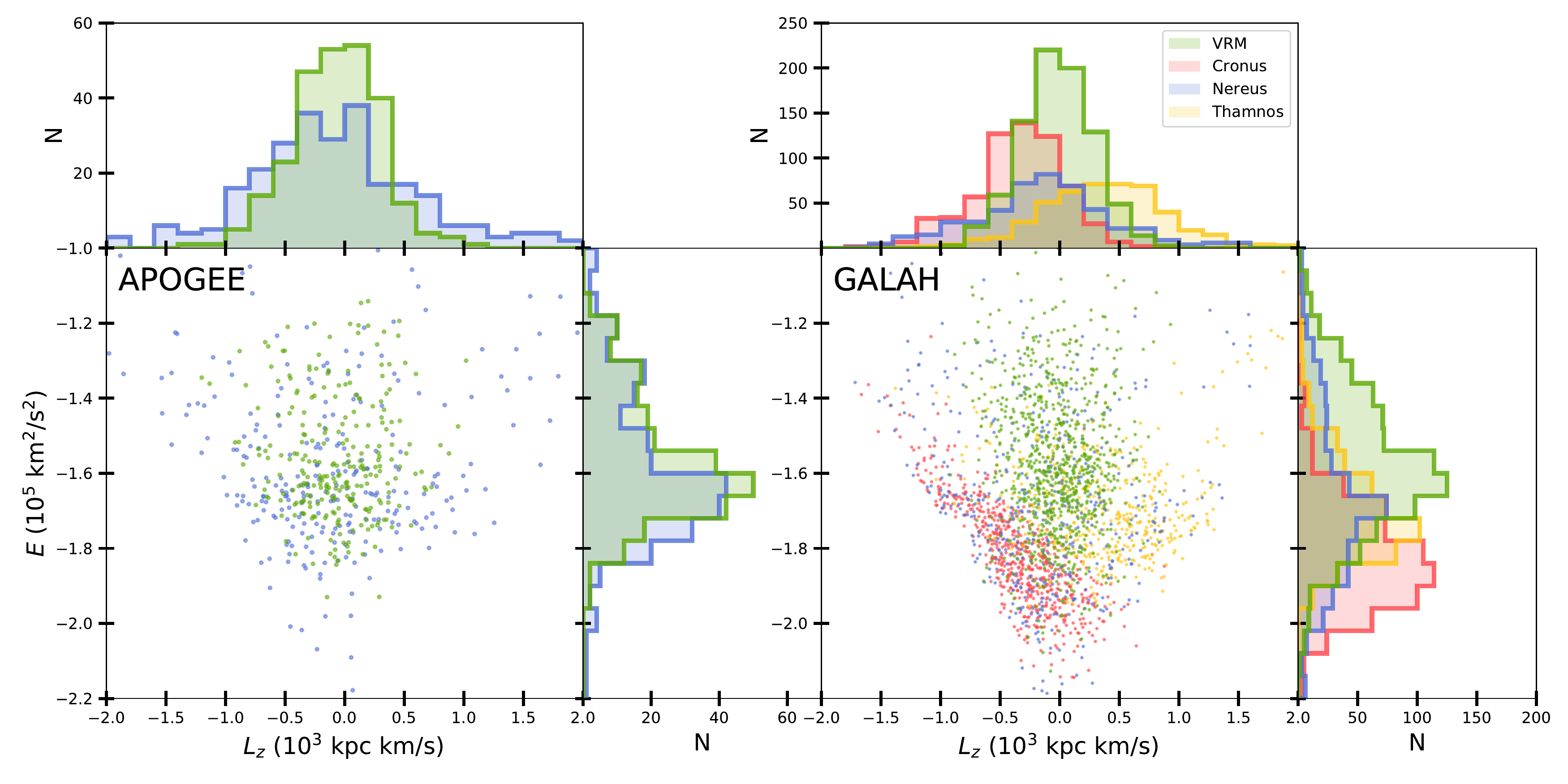}
    \caption{Energy and angular momentum of each GMM component. The left half of the figure contains the APOGEE sample, and the right half of the figure contains the GALAH sample. The different components are shown in green (VRM), red (Cronus), blue (Nereus), and yellow (Thamnos). For each sample, there is a scatter plot containing the $E-L_z$ distributions of the stars, as well as a histogram of the $L_z$ values of each component (top) and a histogram of the energy values of each component (right). In both samples, the VRM does not have a clear net rotation, and Nereus appears to be somewhat prograde. In the GALAH sample, Cronus is very prograde, and Thamnos is very retrograde. In APOGEE, Nereus has a slightly lower energy than the VRM. In GALAH, Cronus has the lowest energy out of the three major local structures, followed by Thamnos, then Nereus, and then the VRM. The VRM stars populate the region of the $E-L_z$ diagram with $L_z\sim0$ that is often attributed to GSE stars, although stars from the other components are also present in this region.}
    \label{fig:e_lz}
\end{figure*}

In addition to the dynamical parameters that are available in the GALAH value-added catalogs, we introduce an additional dynamical parameter, \begin{equation}
r_m = \frac{r_{apo} + r_{peri}}{2},
\end{equation} which has previously been used to characterize possible VRM stars by \cite{ZhaoChen2021}. For structures that predominantly contain stars on radial orbits, this quantity can be more useful than either the apogalacticon or perigalacticon distances on their own; this is because apogalacticon distance is functionally identical to total orbital energy for radial orbits, and perigalacticon distance is usually set by the radial orbit's eccentricity. Additionally, $r_m$ and $z_\textrm{max}$ appear to contain information about orbital resonances with the Galactic bar \citep{Moreno2015,ZhaoChen2021}.

Figure \ref{fig:dynamics} contains the distributions of $v_z$, orbital eccentricity, $z_\textrm{max}$, and $r_m$ for all of the GMM components. In the GALAH data, the VRM has larger orbital eccentricity, $z_\textrm{max}$, and $r_m$ values than the other components. Cronus has the lowest $r_m$ values, which indicates that Cronus stars are typically located closer to the center of the Galaxy than the other three components. In the APOGEE data, the VRM has a slightly larger eccentricity, $r_m$, and $z_{\textrm{max}}$ than Nereus, although the values are also consistent with being the same for both structures. In the GALAH data, Cronus also has a net positive $v_z$, while the VRM stars have a slightly negative $v_z$, and Nereus and Thamnos stars appear to have no net $v_z$ overall. In the APOGEE data, the VRM and Nereus stars both have near zero overall $v_z$. In dynamical equilibrium, the $v_z$ distribution of a population is expected to be symmetric around zero, so the offsets in $v_z$ for the VRM and Cronus stars either indicate selection effects, or that their stars are not in equilibrium.

Figure \ref{fig:e_lz} contains the energy and $L_z$ angular momentum data for each sample, split up into the different GMM components. In the APOGEE sample, the VRM has no clear net rotation, but Nereus has a prograde $L_z$. The energies of the two components are roughly the same, although the VRM appears to have slightly higher energy than Nereus. In the GALAH sample, Cronus has a net prograde rotation, Nereus has a slight prograde rotation, the VRM has a slight retrograde rotation, and Thamnos has a strong retrograde rotation. Cronus has the lowest energy, followed by Thamnos, then Nereus, and the VRM has the highest energy. In both samples, the $L_z$ and $E$ of the local stellar halo components agree with the analysis of \cite{Donlon2022}, provided that the overall $L_z$ shift is taken into account (App. \ref{app:lz_diff}).

The Sequoia structure \citep{Myeong2019} can be seen in Figure \ref{fig:e_lz} at very retrograde $L_z$ and high $\widetilde{E}$. It is not identified as a distinct halo component in either sample; in the APOGEE data, it is assigned to the Nereus component, and in the GALAH data it is assigned to a mixture of the Nereus and Thamnos components. As the Sequoia is a known halo substructure that our algorithm is not able to identify, it may be the case that there are other distinct structures that our procedure is not capable of separating from other components of the local stellar halo. 

\subsection{Chemical Abundances}

\begin{figure*}
    \centering
    \includegraphics[width=0.8\linewidth]{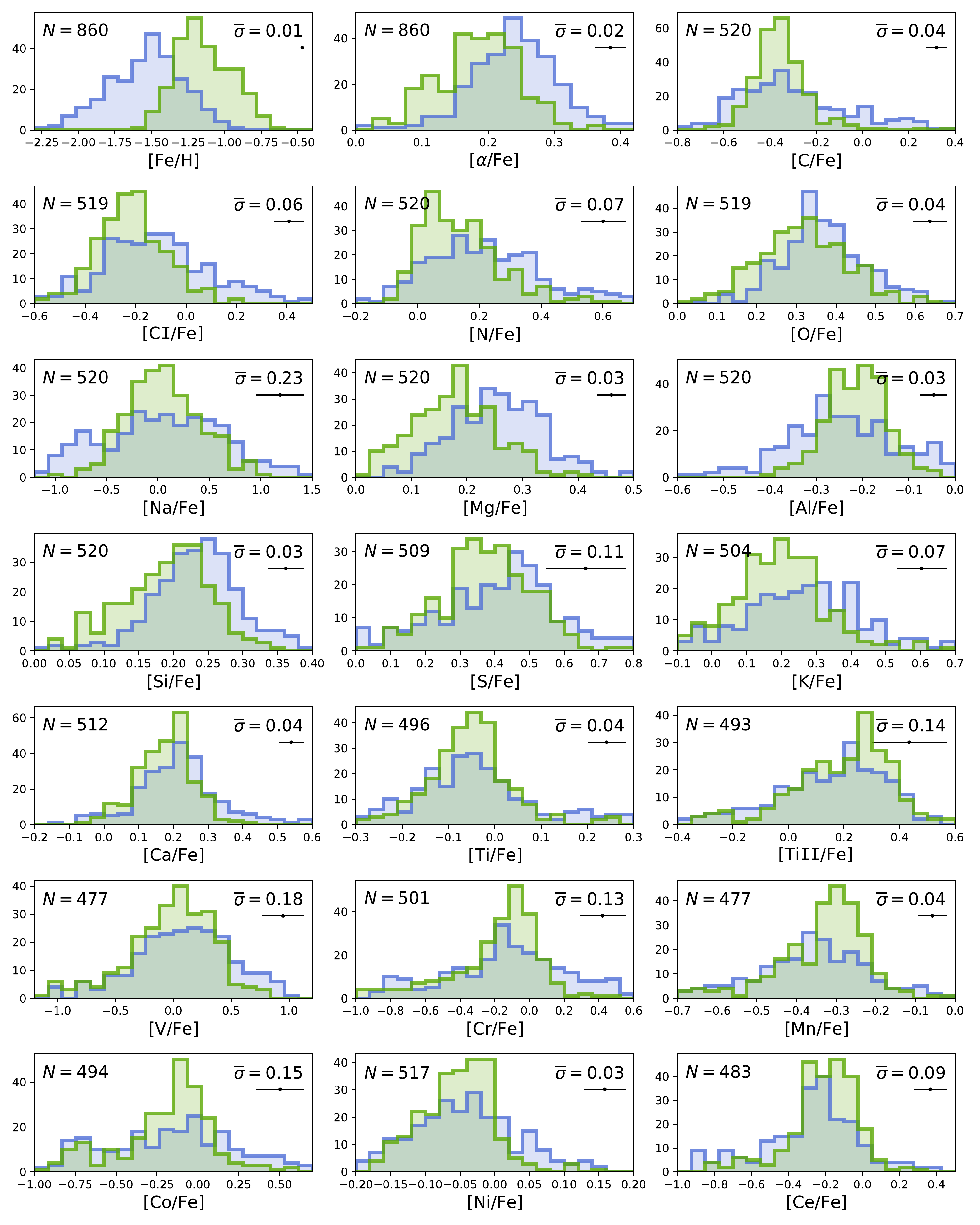}
    \caption{Chemical abundance distributions for the APOGEE stars, separated by color into the components that were determined in Section \ref{sec:model_fit}. The green distribution corresponds to VRM stars, and the blue distribution corresponds to Nereus stars. Each panel corresponds to a given chemical abundance. The number of stars that had a measurement of the given chemical abundance is provided in the top left of each panel. The mean uncertainty in the measurements of the given chemical abundance is provided in the top right of each panel, along with an error bar that illustrates the scale of the mean uncertainty on that panel. Only abundances with over 100 total abundance measurements in our halo sample are shown. There are many abundances (for example, \fe{N}) that were not used to fit our model, but still appear to have a distribution that is different for each GMM component. This could be due to correlations between the APOGEE abundances that we used in the fit and other APOGEE abundances, but it hopefully indicates that we are successfully separating different populations of stars that possess different chemical properties.}
    \label{fig:abundances_apogee}
\end{figure*}

\begin{figure*}
    \centering
    \includegraphics[width=\linewidth]{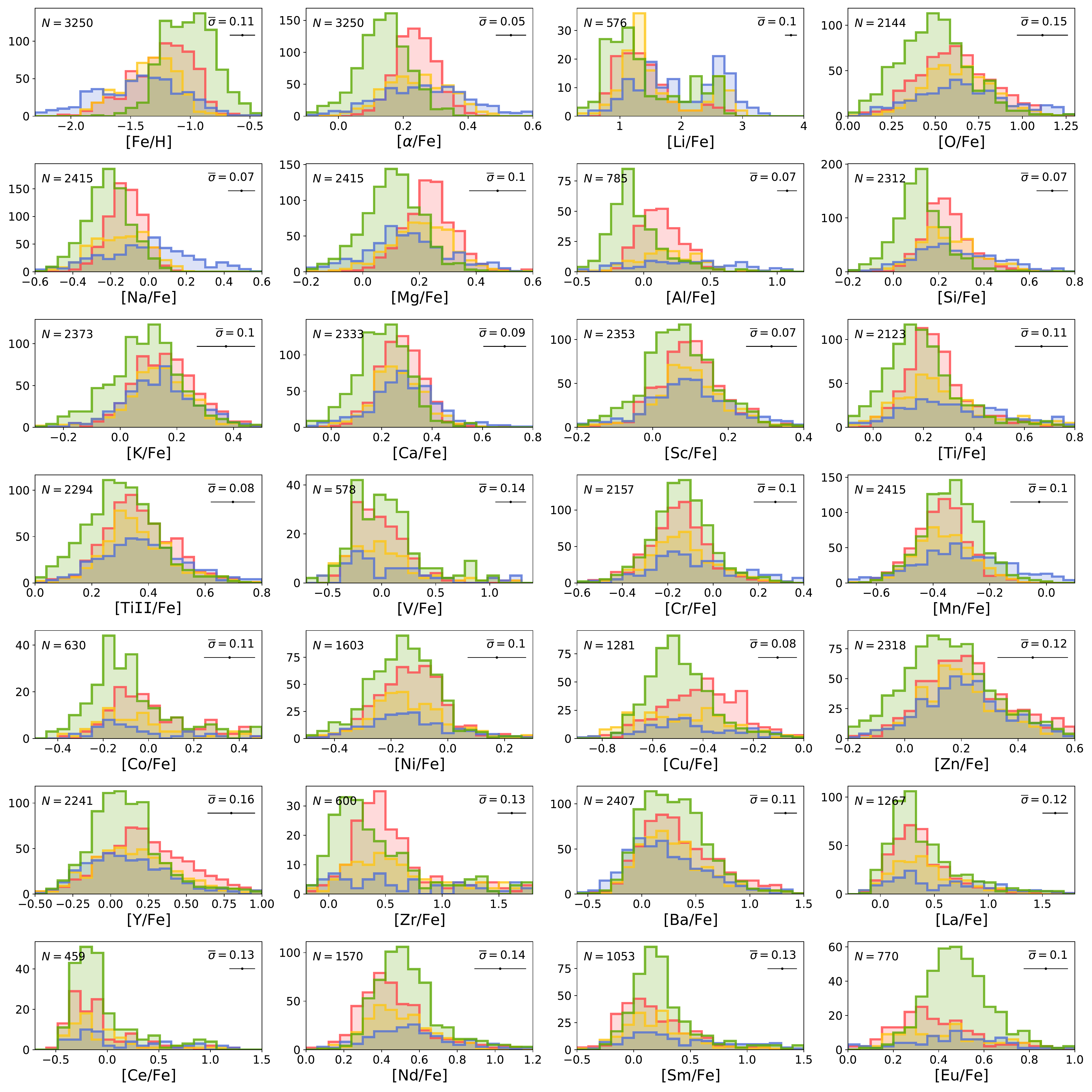}
    \caption{Chemical abundance distributions for the GALAH stars, separated by color into the components that were determined in Section \ref{sec:model_fit}. The green distribution corresponds to VRM stars, the red distribution corresponds to Cronus stars, the blue distribution corresponds to Nereus stars, and the yellow distribution corresponds to Thamnos stars. Each panel corresponds to a given chemical abundance. The number of stars that had a measurement of the given chemical abundance is provided in the top left of each panel. The mean uncertainty in the measurements of the given chemical abundance is provided in the top right of each panel, along with an error bar that illustrates the scale of the mean uncertainty on that panel. Only abundances with over 300 total abundance measurements in our halo sample are shown. As in Figure \ref{fig:abundances_apogee}, the fit components have markedly different distributions for many abundances beyond the abundances that were used to fit the GMM.  }
    \label{fig:abundances_galah}
\end{figure*}

The large number of chemical abundances available to us in the APOGEE and GALAH survey data allow us to obtain detailed chemical abundance distributions for each GMM component. Figure \ref{fig:abundances_apogee} contains histograms of each chemical abundance that was available in the APOGEE data, and Figure \ref{fig:abundances_galah} contains histograms of all the available GALAH chemical abundances. These histograms are split up by component, which shows the differences in chemical composition of each local halo structure.

In both the APOGEE and GALAH data, the VRM is characterized by relatively low abundances of most elements, except for [Mn/Fe] (particularly in the APOGEE data), and possibly \fe{Nd} and \fe{Eu}. This could be due in part to the high [Fe/H] abundance of the VRM, which would appear to correspondingly lower the apparent amount of other elements with respect to Fe. 

In general, Cronus stars have larger \fe{X} chemical abundances than VRM stars. This is especially true for \fe{O}, \fe{Na}, \fe{Mg}, \fe{Al}, \fe{Si}, \fe{Ca}, \fe{Ti}, \fe{TiII}, \fe{Co}, \fe{Cu}, \fe{Y}, and \fe{Zr} compared to VRM stars. As O, Mg, Si, and Ca are \als elements, the high abundances of these elements in Cronus stars is reflected in Cronus stars' high \fe{\al} abundances. 

Cronus' [Fe/H] content is similar to that of the larger classical dwarf galaxies such as the Fornax dSph or Leo I \citep{Kirby2013}. The high \al-element abundances of Cronus stars indicate that the progenitor of Cronus terminated star formation before Type Ia supernovae elements could enrich the gas within the progenitor. In contrast, stars in the Fornax dSph have roughly solar levels of \fe{Mg} and \fe{Si}, which suggests that it sustained multiple epochs of bursty star formation \citep{Hendricks2014}. One way to explain Cronus stars having both high [Fe/H] and high \fe{\al} is if the Cronus progenitor was artificially quenched early on in its star formation history. It is plausible that this quenching could have been due to the Cronus progenitor being accreted by the MW.

In order to generate an [Fe/H] of $\sim$ -1.5 while still maintaining a high \fe{\al} abundance, the progenitor of Cronus was likely more massive than the classical MW dSphs. However, there is a somewhat smaller number of Cronus stars in the local Solar region than VRM stars. This could be due to Cronus' location deeper in the MW's potential well; Cronus stars are preferentially distributed closer to the Galactic Center (e.g. the bottom panel of Figure \ref{fig:dynamics}), whereas VRM stars are distributed throughout the inner halo. It could also be because the VRM debris not phased mixed \citep{Donlon2020}, so there could be a higher density of VRM stars in the local stellar halo than the average density of VRM stars in the MW.

As Mn is primarily produced in Type Ia supernovae \citep{Kobayashi2020}, the large values of \fe{Mn} in VRM stars indicate that the VRM population continued forming stars long enough to experience chemical enrichment via SNe Ia. When a dwarf galaxy is accreted, it is expected that its star formation is quenched; therefore, the VRM was probably only accreted recently, because we expect that it had a lengthy star formation history which was not quenched long ago. In contrast, the large \fe{\al} abundance of Cronus compared to its [Fe/H] abundance indicates that it may have been accreted earlier in the MW's history than the VRM, as it is plausible that its star formation was quenched when it was accreted. The [Fe/H] abundances of VRM stars indicate that the progenitor of the VRM was probably similar in size to the larger classical dSphs, such as Fornax and Leo I. 

Cronus stars have higher abundances of \fe{Y}, and \fe{Zr} and lower abundances of \fe{Nd}, \fe{Sm}, and \fe{Eu} compared to VRM stars. \fe{Y}, and \fe{Zr} are s-process elements, which are thought to be formed primarily in asymptotic giant branch stars \citep{Busso1999}. \fe{Nd}, \fe{Sm}, and \fe{Eu} are r-process elements, which are primarily generated in neutron star mergers \citep{Kobayashi2020}. Interestingly, the VRM and Cronus stars appear to have similar \fe{Ba}, \fe{La}, and \fe{Ce} abundance distributions, which are also neutron-capture elements. The differences in r-process and s-process element abundances in the VRM and Cronus stars likely contain information about the star formation histories of the two structures.

\section{Radial Action \& Chemical Evolution Paths} \label{sec:action_ceps}

\begin{figure}
    \centering
    \includegraphics[width=\linewidth]{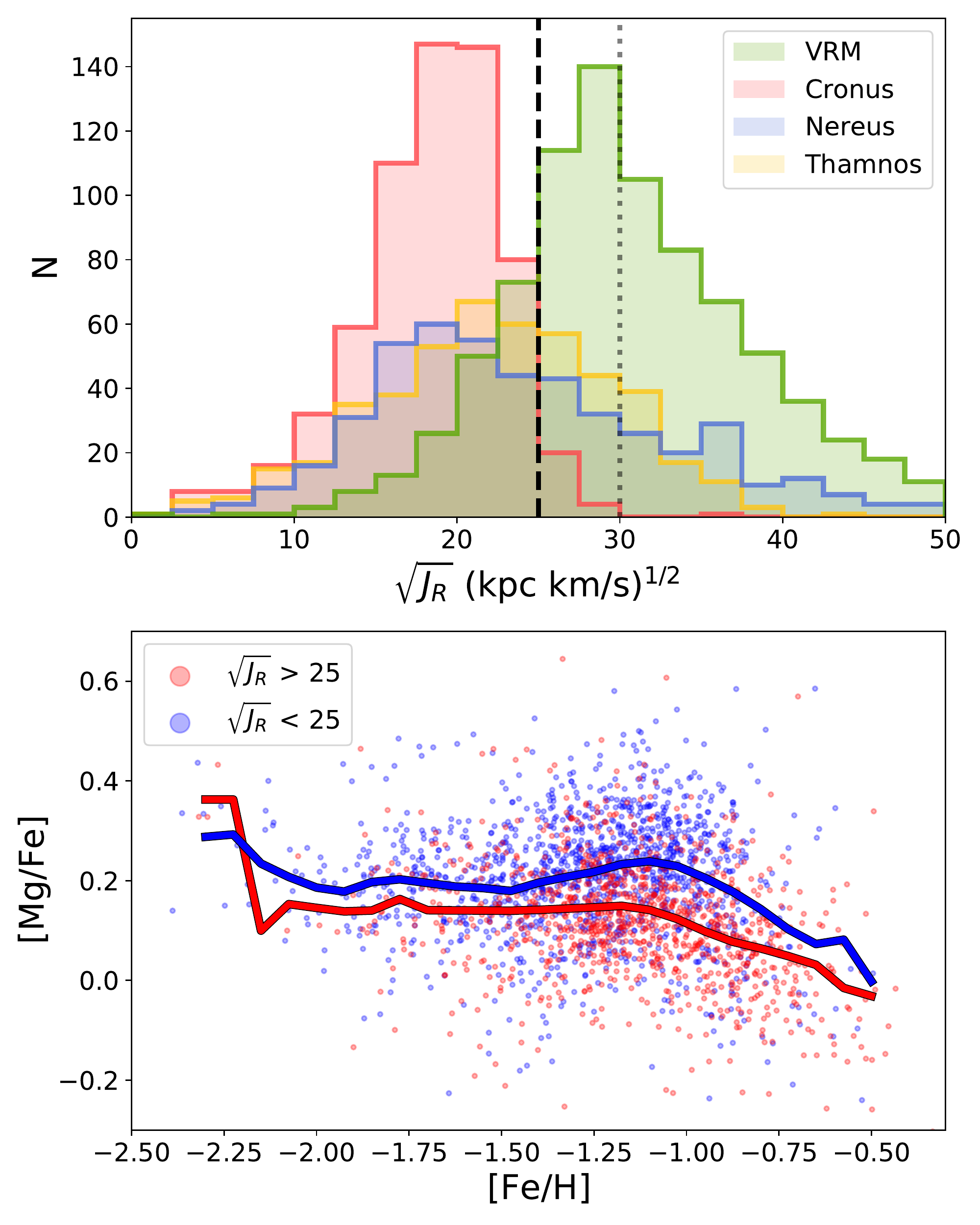}
    \caption{\textit{Top:} Histogram of $\sqrt{J_R}$ for the GALAH GMM components, split up by color. The VRM and Cronus have clearly distinct $\sqrt{J_R}$ distributions, and the Nereus and Thamnos components span a wide range of $\sqrt{J_R}$ values. A cut of $\sqrt{J_R} > 30$ (kpc km s\invnospace)$^{1/2}$ is often used to isolate GSE stars, and is shown by the dotted gray line. A cut of $\sqrt{J_R} > 25$ (kpc km s\invnospace)$^{1/2}$, shown as a dashed black line, does a better job of separating VRM and Cronus stars. However, there are still a number of VRM stars with $\sqrt{J_R}$ smaller than 25 (kpc km s\invnospace)$^{1/2}$, and the Nereus and Thamnos components are present above and below this cut. \textit{Bottom:} \fe{Mg} vs. [Fe/H] for the GALAH stars, split up into stars with $\sqrt{J_R} > 25$ (kpc km s\invnospace)$^{1/2}$ (red, mostly VRM stars) and $\sqrt{J_R} < 25$ (kpc km s\invnospace)$^{1/2}$ (blue, mostly Cronus stars). The blue and red selections each contain roughly half of the stars in the local stellar halo. The mean \fe{Mg} for a given [Fe/H] is plotted for both selections as solid colored lines, which is approximately a chemical evolution path (CEP). The stars with high $\sqrt{J_R}$ have a CEP with lower \fe{Mg} than the stars with low $\sqrt{J_R}$. This is not expected if the local stellar halo is primarily built up from a single merger event, which should have a single CEP. }
    \label{fig:jr_abundances}
\end{figure}

\begin{figure*}
    \centering
    \includegraphics[width=\linewidth]{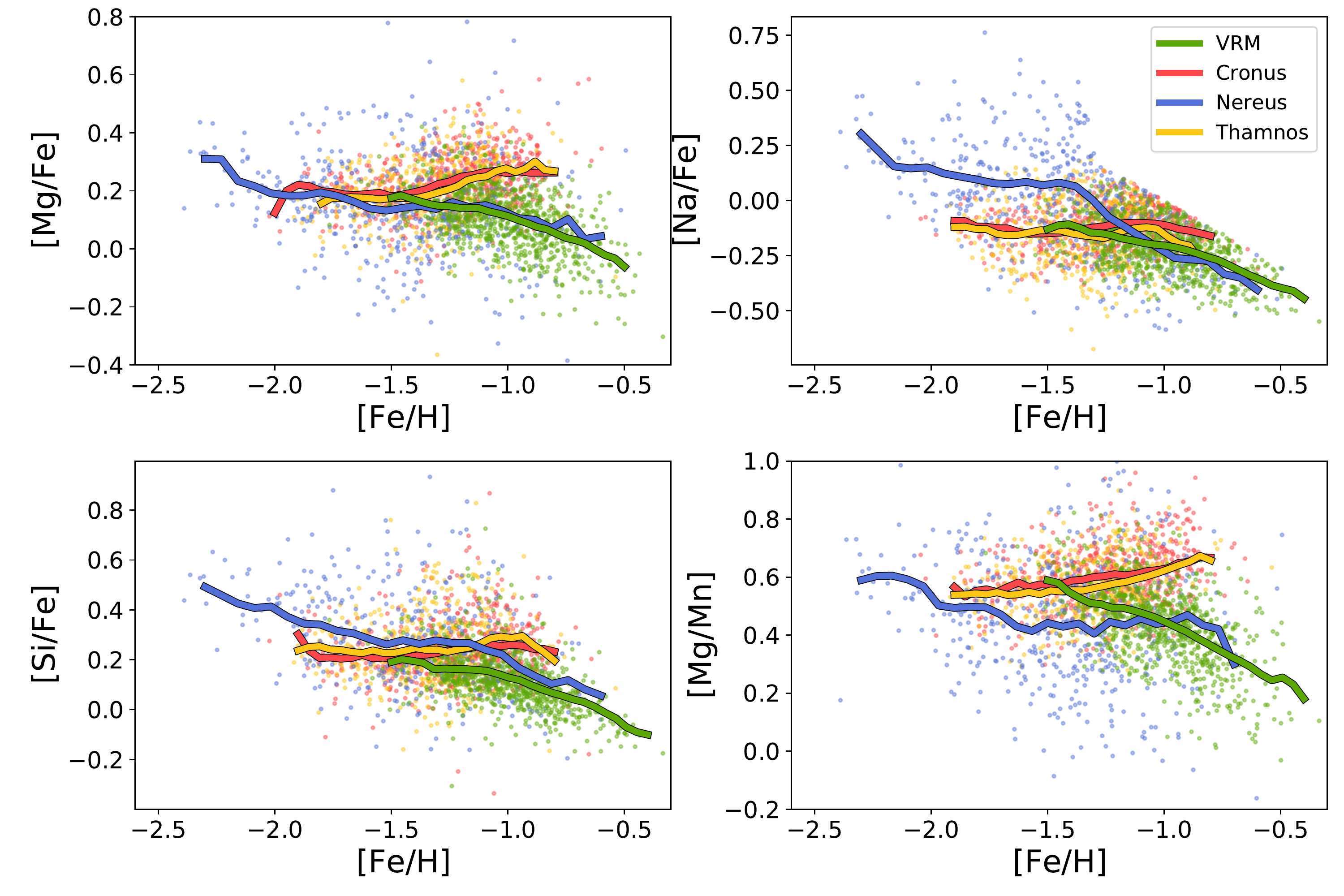}
    \caption{Various abundances vs. [Fe/H] for each GALAH GMM component, which are split up by color. The mean abundance for a given [Fe/H] value is drawn on top of the points for each component, which is approximately equal to a CEP for that component. The CEPs of the four components are all distinct, except for Cronus and Thamnos, which are nearly identical, and may suggest that the two structures are related in some way despite their different kinematics. The top-right panel is missing data due to the chemical cut made to remove the thick disk/Splash stars (Figure \ref{fig:disk_splash}), so the CEPs are artificially skewed to lower \fe{Na} for [Fe/H] $>$ -1.3.}
    \label{fig:ceps}
\end{figure*}

In this section, we use the chemical abundances and actions of stars in order to explore whether (i) radial action cuts produce an unbiased sample of the local stellar halo (specifically the stars that are typically associated with the GSE), and (ii) whether the components we identified in this work are chemically consistent with a single massive dwarf galaxy. We find that the typical radial action cuts used in the literature predominantly select VRM stars, and that the chemical evolution paths of our halo components indicate that their progenitors had different origins, although some of the components may be related to each other. 

\subsection{Radial Action Cuts \& Local Stellar Halo Populations}

In order to further explore the dynamics of our halo star data, we made use of the radial orbital action $J_R$. The radial action can be thought of as the magnitude of a star's orbital oscillation towards or away from the Galactic center; near the Sun, the radial action is essentially equal to the part of a star's energy that comes from its radial motion. As the potential energy of all stars in the local solar region is roughly similar, it follows that $J_R$ is roughly proportional to $v_R^2,$ or equivalently $\sqrt{J_R}\propto|v_R|.$

Within the local stellar halo, GSE debris is characterized as having a large spread of radial velocity, and low spread in rotational velocity \citep{Belokurov2018}. As a result, it will have a large range of $|v_R|$ values, and therefore a large range of $\sqrt{J_R}$ values. Despite this, it is common to isolate GSE debris by only using stars with low $L_z$ and $\sqrt{J_R}$ greater than some value, usually 30 (kpc km s\invnospace)$^{1/2}$ \citep{Feuillet2020,Hasselquist2021,Buder2022}. 

The top panel of Figure \ref{fig:jr_abundances} shows histograms of $\sqrt{J_R}$ for each of the GALAH GMM components. Cronus stars dominate the data with $\sqrt{J_R} < 25$ (kpc km s\invnospace)$^{1/2}$, and VRM stars dominate the data with larger $\sqrt{J_R}$ values. The ``GSE'' stars that are isolated by a cut of $\sqrt{J_R}>30$ (kpc km s\invnospace)$^{1/2}$ are really VRM stars, and are therefore not representative of the chemical and kinematic properties of the local stellar halo as a whole. The often-used cut at $\sqrt{J_R}$ = 30 (kpc km s\invnospace)$^{1/2}$ is higher than the observed transition from Cronus to VRM stars at $\sqrt{J_R}$ = 25 (kpc km s\invnospace)$^{1/2}$. We find it unlikely that using a larger $\sqrt{J_R}$ cut provides any meaningful advantage; the high cut unecessarily removes VRM stars from the sample, and contamination from Nereus and Thamnos stars will occur even in data with $\sqrt{J_R}>40$ (kpc km s\invnospace)$^{1/2}$.

It may not be so surprising that a high-$\sqrt{J_R}$ selection produces a sample that is not representative of the entire local stellar halo. A high-$\sqrt{J_R}$ selection will only select stars that are in the double-lobed velocity structure in the local stellar halo \citep{Donlon2022}, rather than the entire ``sausage'' velocity structure. This selection ignores stars that are located near $(v_r,v_\phi) = (0, 0)$, which make up a substantial fraction of halo stars. If only the stars in the double-lobed velocity structure are considered to be GSE stars, then there must be some other origin for the stars with low $|v_r|$, and the GSE stars cannot make up the majority of the local stellar halo. 

\subsection{Using Radial Action to Isolate Chemical Evolution Paths}

While populations of stars are often expected to clump together in chemical abundances, in reality stars from a forming (dwarf) galaxy will populate a CEP \citep{Andrews2017}. The CEP encodes information about the total mass, star formation history, and chemical abundances of its host galaxy. This is because a given (dwarf) galaxy contains stars from multiple epochs of its star formation, which will each have different chemical abundances due to the overall chemical evolution of the (dwarf) galaxy's gas reserves over time. Typically, CEPs are drawn on plots of an \al-element, such as Mg or Si, against [Fe/H]; this allows one to roughly track the ratio of SNe Ia and SNe II elements in a population over time. 

The bottom panel of Figure \ref{fig:jr_abundances} shows plots of \fe{Mg} vs. [Fe/H] for the GALAH sample, split up into stars with $\sqrt{J_R}$ above and below 25 (kpc km s\invnospace)$^{1/2}$. The CEP for the high $\sqrt{J_R}$ stars sits at a lower \fe{Mg} than the low $\sqrt{J_R}$ stars, which confirms that the $\sqrt{J_R}$ cut is isolating two different chemical populations in the local stellar halo.

Does the existence of two CEPs in the local stellar halo require multiple progenitors? If the progenitor of the GSE was sufficiently massive, then it may have multiple populations of stars (e.g. disk \& halo) that formed in environments with different chemical abundances. This could hypothetically give the populations different CEPs. Then, the halo stars from the GSE progenitor may have been stripped earlier on than the GSE progenitor disk stars, which could potentially give them different kinematics (i.e. Figure \ref{fig:model}). 

There are multiple problems with this hypothesis. Figure 17 of \cite{Hasselquist2021} shows the stars from the Sgr dSph, split up into stars that are still bound to the dwarf, and stars that reside in the Sagittarius Stream tidal tails. It is expected that the present-day stream stars were primarily located in the outer regions of the Sgr dSph, and would have formed in a different chemical environment than the stars within the presently-bound region of the Sgr dSph. However, the stream stars and bound stars are located on the same CEP: the stream stars are located on the left half of the CEP, and the bound stars are located on the right. Based on this, there is no reason to expect that any dwarf galaxy progenitor would have formed multiple different CEPs like we see in the data. One could imagine that the Sgr dSph was just not massive enough to create multiple CEPs, but this is also unlikely, because the LMC data in \cite{Hasselquist2021} does not contain multiple CEPs. However, it is once again easy to explain the existence of multiple CEPs if the local stellar halo contains debris from multiple merger events.

Figure \ref{fig:ceps} shows CEPs of the GALAH GMM components for various elements vs. [Fe/H]. The CEPs of each structure are distinct from one another, except for Cronus and Thamnos, which have extremely similar CEPs for each element. This could suggest that Cronus and Thamnos are related in some way, despite their different dynamics. It is plausible that Cronus and Thamnos were accreted at roughly the same time given their similar energies, and could have been components of a group infall event like the LMC and its satellites today. This idea is supported by the fact that the LMC and SMC, which are known to be accreting via group infall, have extremely similar CEPs \citep{Hasselquist2021}. Additionally, the VRM and Nereus CEPs are not identical, but they are similar to each other. It is also possible that the progenitors of the VRM and Nereus were also associated with a group infall event, given their similar CEPs and energies.

\section{The Nature of Nereus \& Thamnos} \label{sec:nature_nereus}

\begin{figure*}
    \centering
    \includegraphics[width=\linewidth]{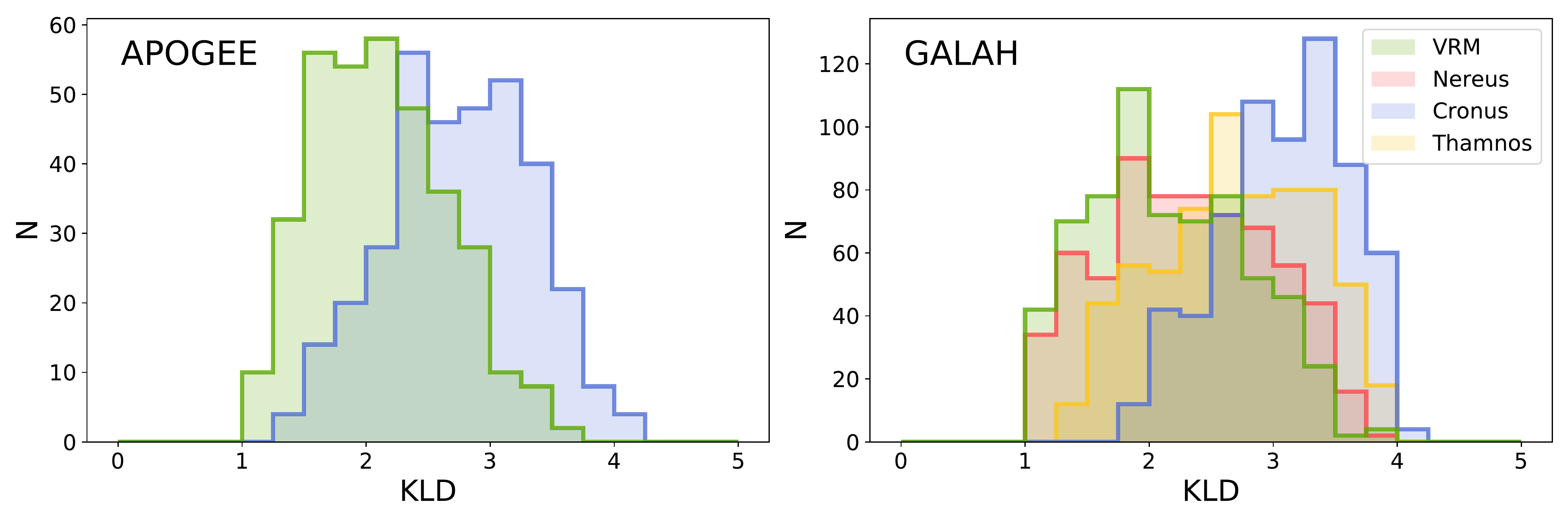}
    \caption{Histograms of KLD values for combinations of all abundances for stars within each halo component, split by survey data. The left panel contains the data from APOGEE, and the right panel contains the data from GALAH. Each halo component is given a different color: the VRM is in green, Cronus is in red, and Nereus is in blue. The KLD score of a distribution indicates how clustered it is; a higher (lower) KLD indicates more (less) clustering. The abundances of Nereus stars have substantially larger KLD values than the VRM and Cronus stars, which indicates a larger amount of clustering within the Nereus stars' abundance distributions. This is possibly explained if Nereus is made up of several minor merger events rather than a single progenitor; each minor merger progenitor would have a slightly different star formation history and environment, which would give each progenitor a specific abundance distribution with small instrinsic scatter. When one sums together the abundance distributions for each minor merger, the result would appear to be a clumpy distribution. }
    \label{fig:kld}
\end{figure*}

\begin{figure*}
    \centering
    \includegraphics[width=\linewidth]{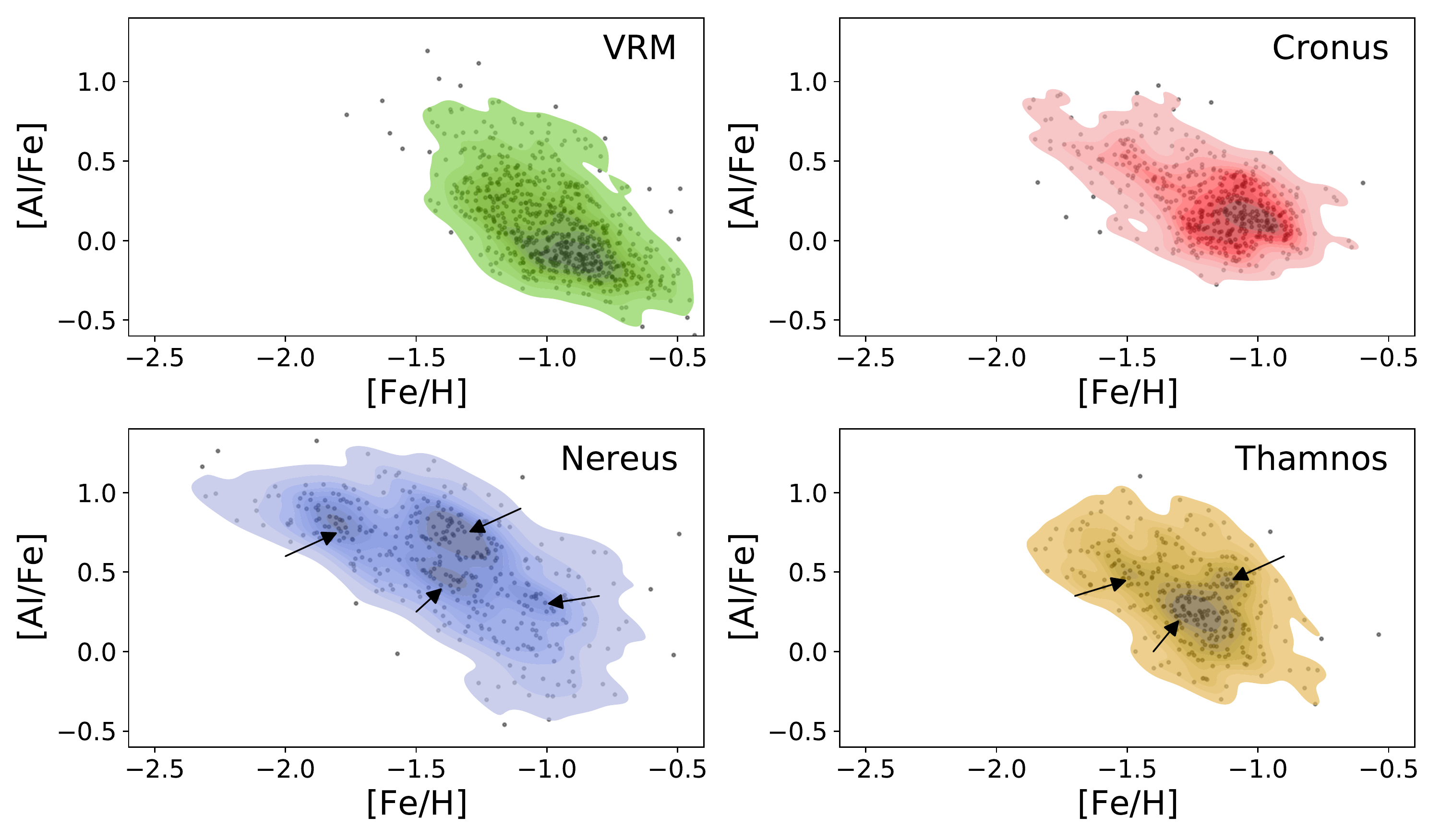}
    \caption{\fe{Al} vs. [Fe/H] for GALAH stars, split up by panel into the GMM halo components. Contours show the relative densities of the stars in each panel. Nereus stars occupy a larger area of the diagram than Cronus or VRM stars. Additionally, Nereus and Thamnos stars clump in many different spots on the \fe{Al}-[Fe/H] plot; some apparent clumps are designated with black arrows. The Cronus and VRM stars do not share this apparent clumpiness, but are instead made up of a single, main cluster of stars. This is visual confirmation of the results of Figure \ref{fig:kld} that Nereus and Thamnos stars are more clumped than VRM and Cronus stars. It is also consistent with the claim that Nereus and Thamnos are likely made up from several progenitors with different abundance distributions. }
    \label{fig:clumpy}
\end{figure*}

In this section we explore the hypothesis that the identified Nereus and Thamnos components are each built up from more than one progenitor. We find that their chemical abundance distributions are more clustered than the chemical abundance distributions of the VRM and Cronus, and that there are visible clumps of stars in the Nereus and Thamnos chemical abundance distributions; these findings are consistent with Nereus and Thamnos being built up from multiple progenitor dwarf galaxies.

The large [Fe/H] abundances of the VRM and Cronus indicate that their progenitors were probably fairly massive, as it is not possible for low-mass dwarf galaxies to maintain star formation for long enough to enrich their gas with Sne Ia elements. In contrast, the lower [Fe/H] and larger \fe{\al} abundances of Nereus and Thamnos suggests that their progenitors were lower mass than the progenitors of the VRM and Cronus. Because the VRM is not phase-mixed \citep{Donlon2020}, and it is possible that Cronus is also in disequilibrium (see discussion of Figure \ref{fig:dynamics}), we cannot simply use the star counts of the local stellar halo to determine the ratios of the components' stellar masses.

One peculiar trend in the data is that Nereus and Thamnos stars have larger dispersion for many chemical abundances than VRM and Cronus stars. This large dispersion could be explained if Nereus and Thamnos are actually composed of the debris from many smaller merger events, rather than a single large progenitor such as the VRM and Cronus. 

If Nereus or Thamnos were built up of smaller merger events, then each individual progenitor would have formed through its own star-formation history in its own environment, and would therefore have its own distinct abundance distribution compared to each of the other small progenitors. Each progenitor's abundance distribution would have a small intrinsic scatter compared to the overall population because of the progenitor's small mass. This would cause the abundance distributions of Nereus/Thamnos to contain many small clumps, each one corresponding to an individual progenitor. Conversely, if Nereus/Thamnos were formed from a single massive progenitor, then one would expect a single large distribution without any smaller abundance clumps. 

One way of evaluating whether any of our halo components contain deeper substructure is by quantitatively measuring the amount of clustering, or ``clumpiness'', of the stellar abundances within each component. To this end we utilize a form of Kullback-Leibler Divergence \citep[KLD,][]{Kullback1951,Fabricius2021}, which is defined as \begin{equation}
\textrm{KLD} = \int_A P(\mathbf{x}) \log \left( \frac{P(\bf{x})}{Q(\bf{x})}\right)\;\textrm{d}^2x,
\end{equation} where $P(\mathbf{x})$ is the 2-dimensional joint probability density function, and $Q(\mathbf{x}) = \Pi_i P_i(x_i),$ which is the product of the marginalized 1-dimensional probability density functions. The KLD is larger for more clustered distributions and smaller for less clustered distributions, which allows us to measure how clustered a given 2-dimensional distribution is. 

For each halo component, we numerically computed the KLD for each combination of abundances within the APOGEE and GALAH datasets. The results of this are shown in Figure \ref{fig:kld}; in the APOGEE data, the KLD of Nereus abundance distributions are much larger than the KLD for VRM stars, and in the GALAH data the mean KLD of Nereus is larger than the KLD of Thamnos, which is larger than the KLDs for Cornus and the VRM. This indicates that the abundance distributions of Nereus and Thamnos stars are more clustered than the abundance distributions of VRM and Cronus stars. This supports our hypothesis that Nereus and Thamnos are actually built up of many small accretion events, rather than a merger event with a single progenitor. Because Nereus has an even higher KLD than Thamnos, it is reasonable to believe that it is built up from more progenitors than Thamnos.

This clumpiness is apparent in abundance diagrams; one good example is \fe{Al} vs. [Fe/H], which is shown in Figure \ref{fig:clumpy}. While the majority of VRM and Cronus stars cluster into a single group, the Nereus and Thamnos stars clump into several apparent substructures. It is easiest to explain the multiple peaks in the chemical abundance distributions of Nereus and Thamnos stars if they are built up from the material of multiple progenitors that formed independently.

These results are in tentative agreement with the results of \cite{Koppelman2019b}, who found that Thamnos has two distinct populations. While they did not find any evidence that the different populations in Thamnos were required to be from different progenitors, we believe that this new evidence warrants further consideration of the idea that Thamnos is not built up from a single merger event.

Determining the statistically significant number of abundance clumps within Nereus and Thamnos, the relative size of each clump, and if/how these clumps correspond to different progenitors will require detailed analysis. These questions have important implications for the accretion history of the MW, including potentially constraining the number of small merger events with radial orbits that happen at various times, the average (stellar) mass of these minor merger events, and at what times these minor mergers happen.

\section{Comparisons With Other Literature} \label{sec:lit}


In this section we compare our conclusions with those of other literature. Importantly, we show that we are able to explain the results of \cite{NissenSchuster2010} as a combination of VRM and Cronus stars, rather than being separate populations of accreted and in-situ stars.

\subsection{Nissen \& Schuster (2010)}

One major work that analyzed potential populations of halo stars was \cite{NissenSchuster2010} (hereafter ``NS10''), who used the kinematics and \fe{Mg}, \fe{Ni}, and \fe{Na} abundances of halo stars in order to show that there were two distinct populations of stars in the local stellar halo. They found that there was a population of stars with low \fe{\al} abundances and net retrograde motion, as well as a chemically-distinct population of stars with high \fe{\al} abundances and net prograde motion. 

These two halo populations have since been interpreted as accreted (low-\al) and ``in-situ'' (high-\al) populations within the halo \citep{Buder2022,Belokurov2022}, and the NS10 low-\als stars have even been used as a guide for selecting stars that belong to the GSE \citep{Buder2022}. 

However, we offer an alternative interpretation of the NS10 data. First, we remove all NS10 stars that were classified as belonging to the thick disk, as well as any NS10 stars that are above the Splash/thick disk chemical cut given in Equation \ref{eq:galah_disk_cut}. Then, instead of splitting the NS10 stars into a high and low \fe{\al} population, we split the stars based on whether they are located above or below the line \begin{equation} \label{eq:ns_cut}
    [\textrm{Ni/Fe}] = 1.05\;[\textrm{Na/Fe}] + 0.08.
\end{equation}

\begin{figure*}
    \centering
    \includegraphics[width=\linewidth]{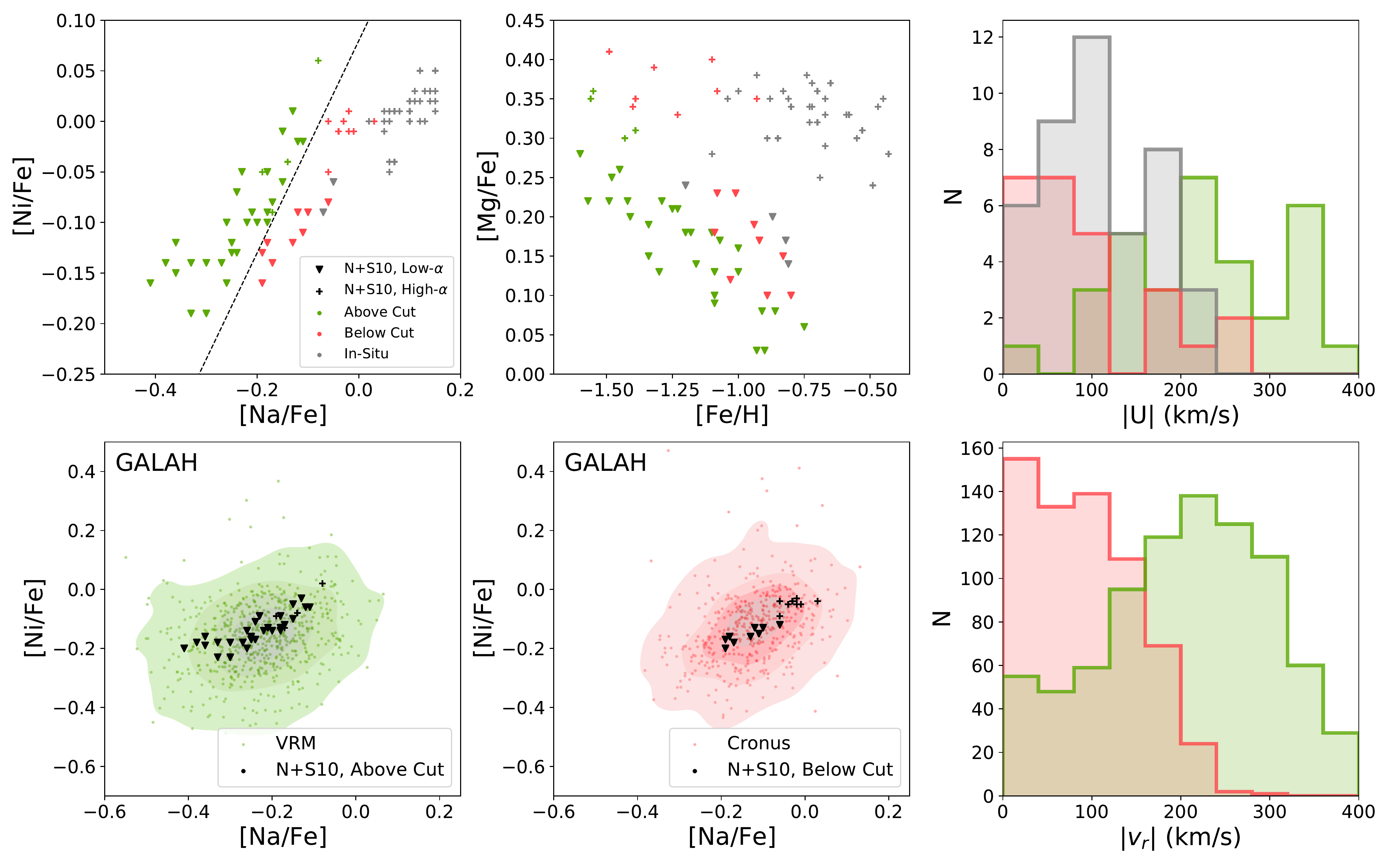}
    \caption{Multiple chemodynamic populations present within the NS10 data. The top row shows the data from NS10, split up into the high [\al/Fe] (crosses) and low [\al/Fe] (triangles) stars as per the analysis of NS10. We remove all in-situ stars based on their [Na/Fe] \& [Fe/H], and then we separate these stars as sitting above (green) or below (red) the line given in Equation \ref{eq:ns_cut}, as shown in the top left panel. This cut separates the stars into high and low chemical evolution paths (top middle panel), and the two selections separate out into stars with high and low magnitudes of Galactocentric $v_x$ velocity ($|U|$, top right panel). In the solar neighborhood, $U\approx v_R.$ The bottom row shows a comparison of the NS10 data with the GMM halo components in the GALAH data. The bottom left panel shows [Ni/Fe] vs. [Na/Fe] for the NS10 stars above the dashed line in the top left panel, compared to the GALAH VRM stars with [Fe/H] $>$ -1.6; the bottom middle panel shows the same for the NS10 stars below the dashed line in the top left panel, compared to the GALAH Cronus stars with [Fe/H] $>$ -1.6. The bottom right panel shows a histogram of $|v_r|$ for the GALAH VRM and Cronus stars with [Fe/H] $>$ -1.6. When the NS10 stars are split based on the cut in Equation \ref{eq:ns_cut}, the resulting two selections of stars are consistent with the VRM and Cronus debris. We do not see Nereus or Thamnos stars in this data, because the [Fe/H] $>$ -1.6 cut in NS10 removes the majority of the Nereus and Thamnos stars from their dataset.}
    \label{fig:nissen_schuster}
\end{figure*}

This separation is shown in the top left panel of Figure \ref{fig:nissen_schuster}. It is apparent in the top middle panel of Figure \ref{fig:nissen_schuster} that these two separate populations appear to trace out two distinct chemical evolution paths. These two CEPs are stacked one on top of the other, which is reminiscent of the bottom panel of Figure \ref{fig:jr_abundances}. This connection is confirmed in the top right panel of Figure \ref{fig:nissen_schuster}, where the two CEPs have different $|v_R| \propto \sqrt{J_R}$ distributions.

These two selections are consistent with the VRM and Cronus stars, as we showed earlier that the VRM and Cronus stars are split in $\sqrt{J_R}.$ The similarity between the two CEPs in the NS10 data and the GALAH VRM and Cronus components is shown in the bottom row of Figure \ref{fig:nissen_schuster}. The \fe{Ni} abundances of NS10 stars in the bottom row of Figure \ref{fig:nissen_schuster} are adjusted by -0.04 dex, to account for the mean offset between NS10 abundances and GALAH data \citep{Buder2022}. There does not appear to be a corresponding group of Nereus or Thamnos stars in NS10, because NS10 only collected data for stars with [Fe/H] $>$ -1.6, and Nereus and Thamnos stars have low [Fe/H].

\subsection{Hierarchical Clustering}

It is common to identify substructure using algorithms which take observed stellar data and assign individual stars to statistically significant groups. These hierarchical clustering algorithms are fundamentally different from typical GMM clustering procedures because they assume that a background distribution exists. Sometimes the background distribution is determined before the fitting process, or it can be determined afterwards as the sum of all stars that did not belong to statistically significant groups. Conversely, in GMM fitting, it is not guaranteed that the fit model will have any components that can be interpreted as ``background''. While some stars may not belong to a statistically significant group in hierarchical clustering, GMM fitting typically assigns every star to at least one component.

\cite{Dodd2022} uses a hierarchical clustering algorithm to identify substructure in the local stellar halo in \textit{Gaia} DR3, LAMOST, and APOGEE data. They find many clusters that are consistent with structures such as the GSE, Thamnos, and the Sequoia. \cite{Dodd2022} identify a structure that they call ``ED-1,'' which appears kinematically and chemically similar to Cronus. 

\cite{Ou2022} also use a hierarchical clustering algorithm on \textit{Gaia} DR3 data cross-matched with LAMOST and APOGEE. They identify several clusters that they claim correspond to the GSE structure, although some of these clusters have high energies, while others have low energies. It is plausible that the high-enegry clusters contain VRM stars while the low-energy clusters contain Cronus stars, and that an energy split is seen in \cite{Ou2022} because their clustering algorithm is separating the two structures. 

\subsection{Other Literature}

\begin{figure}
    \centering
    \includegraphics[width=\linewidth]{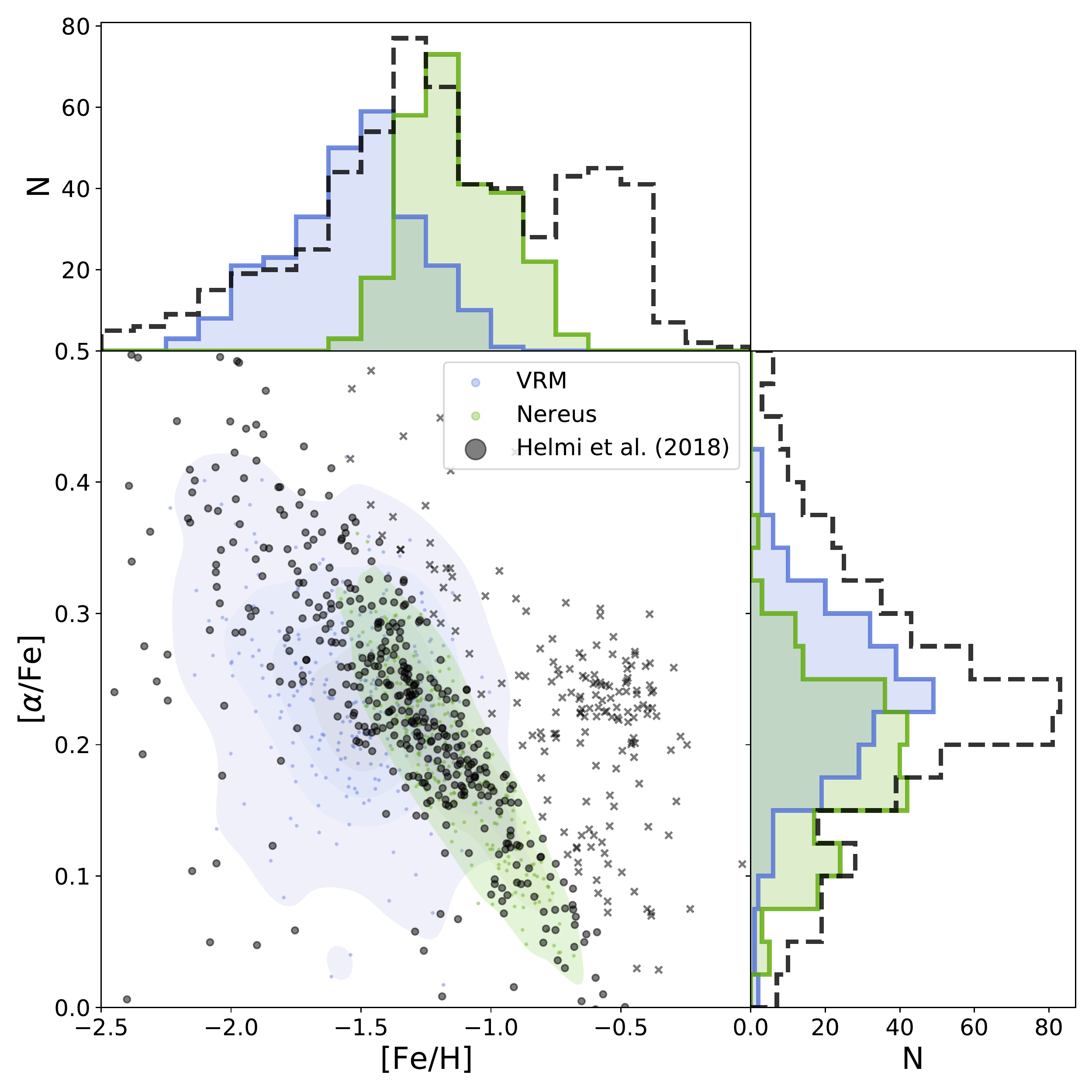}
    \caption{Abundance corner plot of APOGEE stars from this work (VRM in green, Nereus in blue) compared to the \cite{Helmi2018} identification of \textit{Gaia}-Enceladus stars in black. The black circles correspond to their final dataset after a linear \fe{\al}-[Fe/H] cut, and the black crosses indicate their dataset before this cut. Overall, there is good agreement between the H+18 dataset and our data. However, the H+18 data does not include a number of Nereus stars with low \fe{\al} around [Fe/H] $=$ -1.7; this may be due to having more APOGEE data available to us than was available to H+18.}
    \label{fig:helmi}
\end{figure}

\cite{Helmi2018} (hereafter H+18) was one of the two papers that are credited with the discovery of the GSE structure. H+18 provided the first \fe{\al}-[Fe/H] diagram of APOGEE GSE stars in the local solar neighborhood. Figure \ref{fig:helmi} shows a comparison of the H+18 data with our APOGEE GMM halo components. Overall, the H+18 data appears very similar to our VRM component, except that their stars extend up to higher \fe{\al} and lower [Fe/H] than the VRM. The population of Nereus stars with -2.0 $<$ [Fe/H] $<$ -1.5 and 0.15 $<$ \fe{\al} $<$ 0.3 are mostly missing from the H+18 data, although the very \fe{\al}-rich Nereus stars are present in the data. These differences could be due to more APOGEE data being released since H+18, or it could be due to the more restrictive kinematic cuts made by H+18. 

\cite{Das2020} found a ``blob'' of stars in APOGEE data, which they claimed was composed of accreted material based on their chemical abundances. This blob has been associated with the GSE in later works \citep[for example, ][]{Buder2022}, and the chemical abundances of this blob are consistent with that of the stellar halo components identified in this work. \cite{Das2020} found that stars that are located towards the bottom right of a \fe{Mg}-[Fe/H] plot (where we expect most of our VRM stars to be located) have smaller ages on average than stars up and to the left on a \fe{Mg}-[Fe/H] plot (where Cronus, Nereus, and Thamnos stars are found). This supports our claim that the progenitor of the VRM was accreted recently, and was able to support an extensive period of star formation. 

\cite{Buder2022} also uses a GMM procedure to identify populations of stars in GALAH data; however, their GMM models only fit using chemical abundances, without any kinematic information. They find that only two of these components separate meaningfully from the others: an accreted component they identify as the GSE, and an ``in-situ'' component with high \fe{Na} abundances. This accreted component has chemical abundances similar to those of the VRM, Cronus, and Nereus. Interestingly, in one of their GMM fits (panel ``e'' in their Figure 6), their GMM appears to fit two distinct low-[Fe/H] and high-[Fe/H] components to the accreted stars. The low-[Fe/H] component probably corresponds to Nereus/Thamnos, while the high-[Fe/H] component probably contains the VRM and Cronus. In general, the \cite{Buder2022} GMM procedure is not able to separate the VRM, Cronus, Nereus, and Thamnos components from one another (though the stated goal of \citealt{Buder2022} was to identify accreted stars within the halo, and not to identify substructures within those accreted stars). This indicates that the orbital dynamics of the local halo stars provide essential information for disentangling the multiple components of the local stellar halo.

\cite{Buder2022} discusses the differences between the typical dynamical cut that is used to isolate GSE stars, and a chemical cut that was used to isolate likely GSE stars. They find that there is some overlap between the two selections, but that the dynamical cut and the chemical cut select different samples of stars. \cite{Buder2022} interpret that this is because the overlap between the dynamical cut and the chemical cut provides a sample of GSE stars with low contamination and relatively low completeness, whereas the other two cuts have high contamination but higher completeness. We argue that these cuts are in reality selecting different structures within halo stars: the chemical cuts select the majority of the accreted material in the halo, but the kinematic cuts are specifically isolating high-$\sqrt{J_R}$ stars (which primarily belong to the VRM debris), so it makes sense that the two selections do not produce identical samples. 

\cite{Han2022b} find that the MW's stellar halo is doubly broken, with one stellar break located at $r$ = 12 kpc, and another located at $r$ = 28 kpc. They interpret these distances as pericenters in the orbit of the GSE progenitor, where large amounts of material would have been stripped from the dwarf galaxy. We offer an alternative interpretation of this result; the inner stellar break could be due to apocenter pile-up of Cronus material, while the outer stellar break could be due to apocenter pile-up of VRM stars. This is consistent with our results that Cronus stars have lower energy than VRM stars, and that the mean orbital distance of Cronus stars is smaller than that of VRM stars. Thamnos and Nereus material likely came from multiple lower-mass progenitors, so it is unlikely that they would produce strong stellar breaks in the halo. 

\section{Conclusions} \label{sec:conclusion}

In the last few years, there has been an explosion of literature surrounding the identification of the ``last major merger,'' which has since been named the \textit{Gaia}-Enceladus or Gaia Sausage (GSE) merger event. This literature is based on the central idea that a galaxy with a sizeable mass compared to the proto-MW collided with our Galaxy between 6 and 10 Gyr ago, which had many profound effects on the MW that can still be seen today. 

We argue two points in this work, which offer an alternative viewpoint to the GSE literature: (i) A single massive merger event at early times is not able to explain the observed chemodynamics of the local stellar halo stars, and (ii) The local stellar halo stars can be separated into (at least) four components, which are the Virgo Radial Merger (VRM), Nereus, Cronus, and Thamnos. We claim that all of the phenomena laid out in this paper can be explained by the scenario where many dwarf galaxies were accreted onto the Milky Way on more-or-less radial orbits over the Galaxy's lifetime, and are now mostly phased-mixed within the inner halo (except for the VRM).

In order to analyze local stellar halo stars, we removed the thick disk/Splash stars from our sample via a chemical abundance cut. These thick disk/Splash stars appear shifted to lower [Fe/H] and higher $v_\phi$ compared to the selection box in \cite{Belokurov2020}. This group of high-$v_\phi$ Splash stars does not appear to bend into the typical disk distribution because we made kinematic cuts that removed stars with disk kinematics from the data. 

Our claim that a single, massive, early merger is inconsistent with observations is based on a number of chemodynamic trends within APOGEE and GALAH data. We create qualitative models for the infall of a single massive, ancient merger event with an internal metallicity gradient. In this scenario, the accreted stars with lower-[Fe/H] debris would have larger energy and $L_z$ with respect to the host galaxy than the accreted high-[Fe/H] stars. In both surveys, we show that halo stars with lower (higher) energy have chemical abundances that trend the opposite way compared to our single merger model. Additionally, the $L_z$-abundance trends of the observed data would require a prograde orbit of the GSE progenitor, which is not consistent with the results of other studies. These trends cannot be explained by our model of the accretion of a single massive, ancient dwarf galaxy such as the GSE. However, it is simple to explain these chemodynamic trends if the local stellar halo actually contains debris from multiple merger events. 

We show that stars with high $\sqrt{J_R}$ and stars with low $\sqrt{J_R}$ lie on different chemical evolution paths. Additionally, the CEPs of the four identified halo components are distinct, except for Cronus and Thamnos, which may be related in some way. The existence of multiple chemical evolution paths, each with different kinematics, is not consistent with a single, massive, early merger event. Even if one assumes that the GSE contributes the majority of stars in the local stellar halo, this result indicates that the kinematic selection of $\sqrt{J_R}>30$ (kpc km s\invnospace)$^{1/2}$, which is often used to select GSE stars, does not produce a representative sample of the accreted stars near the Sun. We are able to explain the existence of several chemical evolution paths by the existence of debris from multiple merger events in the local stellar halo.  

Our claim that the stars in the local stellar halo can be separated into (at least) four components is based on a Bayesian Gaussian mixture model regression algorithm, which finds that the best model fit for the GALAH local stellar halo stars contains four components. The GMM algorithm preferred two components for the APOGEE data due to the different footprints of the two surveys, which limits the energy range of APOGEE stars. We interpret these components as inner halo substructure, and each component corresponds to a particular merger event (or set of merger events). The identified substructures are summarized here: \begin{itemize}
\item \textit{VRM}: A high-[Fe/H], low-\fe{\al} substructure with large energy and a small retrograde rotation. The high [Fe/H], low \fe{\al}, and large energy of this structure suggest that it was likely accreted recently. The VRM is closest to the typical characterization of the GSE, as it has high orbital eccentricity, creates a double-lobed $v_r$-$v_\phi$ velocity structure in the solar neighborhood, and has large $\sqrt{J_R}$. 

\item \textit{Cronus}: A high-[Fe/H], high-\fe{\al} substructure with low energy and prograde rotation. The large \fe{\al} values of the Cronus stars indicate an early time of accretion, which is also consistent with Cronus stars' low energy. 

\item \textit{Nereus}: A low-[Fe/H], high-\fe{\al} substructure with intermediate energy, and little-to-no net rotation. Nereus stars form several clumps in abundance space, which indicates that Nereus is probably made up of material from several smaller progenitors that formed independently and then fell in at relatively early times. 

\item \textit{Thamnos}: This component has a relatively low energy, relatively low [Fe/H], and contains stars on strongly retrograde orbits. Thamnos stars' chemical abundances are also clumpy and have a wide dispersion, which is in agreement with the idea that Thamnos came from more than one progenitor dwarf galaxy. 
\end{itemize}

We compare our results to previous analyses of halo stars. In general, these studies assume that all accreted stars arise from a single structure. However, the data from \cite{NissenSchuster2010} is remarkably consistent with our multiple-progenitor picture when split into chemical evolution paths rather than high and low \fe{\al}. 

Some of the typical cuts made on the halo to select GSE stars, such as selecting all stars with orbital eccentricity $>$ 0.8, are really selecting a combination of VRM, Cronus, and Nereus stars. Conversely, some selections, such as $\sqrt{J_R}>30$ (kpc km s\invnospace)$^{-1/2}$, are preferentially selecting for only one of these components. Treating these selections as identical makes sense when one assumes that the entire local stellar halo is composed of debris from a single merger event, but in reality these various selections correspond to different mixtures of substructures within the local stellar halo, which can result in different observed properties depending on how the data was selected. 

The Milky Way's stellar halo was originally considered to be smooth in density; now, we know that the stellar halo contains a multitude of tidal streams, which provide evidence of many accretion events over a large span of time. It makes sense that the radial portion of the stellar halo would be formed in a similar way, with many accretion events over the course of the Milky Way's history, rather than a single merger event depositing the majority of the material in the inner halo. The picture of a single ancient, massive merger event dominating the MW's inner halo is no longer sufficient to explain the wealth of chemodynamic substructure near the Sun, and the figurative pendulum must swing back to a view where the local stellar halo was built up over time from multiple merger events. 

\acknowledgments

This work was supported by NSF grant AST 19-08653; contributions made by Manit Limlamai; and the NASA/NY Space Grant.

This work has made use of data from the European Space Agency (ESA) mission {\it Gaia} (\url{https://www.cosmos.esa.int/gaia}), processed by the {\it Gaia} Data Processing and Analysis Consortium (DPAC, \url{https://www.cosmos.esa.int/web/gaia/dpac/consortium}). Funding for the DPAC has been provided by national institutions, in particular the institutions participating in the {\it Gaia} Multilateral Agreement.

Funding for the Sloan Digital Sky Survey IV has been provided by the Alfred P. Sloan Foundation, the U.S. Department of Energy Office of Science, and the Participating Institutions. SDSS-IV acknowledges support and resources from the Center for High Performance Computing  at the University of Utah. The SDSS website is \url{www.sdss.org}. SDSS-IV is managed by the Astrophysical Research Consortium for the Participating Institutions of the SDSS Collaboration including the Brazilian Participation Group, the Carnegie Institution for Science, Carnegie Mellon University, Center for Astrophysics | Harvard \& Smithsonian, the Chilean Participation Group, the French Participation Group, Instituto de Astrof\'isica de Canarias, The Johns Hopkins University, Kavli Institute for the Physics and Mathematics of the Universe (IPMU) / University of Tokyo, the Korean Participation Group, Lawrence Berkeley National Laboratory, Leibniz Institut f\"ur Astrophysik Potsdam (AIP),  Max-Planck-Institut f\"ur Astronomie (MPIA Heidelberg), Max-Planck-Institut f\"ur Astrophysik (MPA Garching), Max-Planck-Institut f\"ur Extraterrestrische Physik (MPE), National Astronomical Observatories of China, New Mexico State University, New York University, University of Notre Dame, Observat\'ario Nacional / MCTI, The Ohio State University, Pennsylvania State University, Shanghai Astronomical Observatory, United Kingdom Participation Group, Universidad Nacional Aut\'onoma de M\'exico, University of Arizona, University of Colorado Boulder, University of Oxford, University of Portsmouth, University of Utah, University of Virginia, University of Washington, University of Wisconsin, Vanderbilt University, and Yale University.

This work made use of the Third Data Release of the GALAH Survey \citep{Buder2021}. The GALAH Survey is based on data acquired through the Australian Astronomical Observatory, under programs: A/2013B/13 (The GALAH pilot survey); A/2014A/25, A/2015A/19, A2017A/18 (The GALAH survey phase 1); A2018A/18 (Open clusters with HERMES); A2019A/1 (Hierarchical star formation in Ori OB1); A2019A/15 (The GALAH survey phase 2); A/2015B/19, A/2016A/22, A/2016B/10, A/2017B/16, A/2018B/15 (The HERMES-TESS program); and A/2015A/3, A/2015B/1, A/2015B/19, A/2016A/22, A/2016B/12, A/2017A/14 (The HERMES K2-follow-up program). We acknowledge the traditional owners of the land on which the AAT stands, the Gamilaraay people, and pay our respects to elders past and present. This paper includes data that has been provided by AAO Data Central (\url{datacentral.org.au}).

\software{gaiadr3\_zeropoint \citep{Lindegren2021}, 
McMillan2017 \citep{McMillan2017}, 
galpy \citep{Bovy2015}, 
scikit-learn \citep{scikit-learn}, 
astropy \citep{Astropy1,Astropy2}}

\bibliographystyle{aasjournal}
\bibliography{references.bib}

%
%
%

\appendix 
\section{Variation of $L_z$ Based on the Chosen Kinematic Frame} \label{app:lz_diff}

The $z$-component of angular momentum can be computed as \begin{equation}
    L_z = x v_y - y v_x.
\end{equation} The calculated Cartesian Galactocentric positions ($x,y$) and velocities ($v_x,v_y$) of each star are functions of distance from the Sun, and are also dependent on the choice of the Sun's phase space coordinates. Therefore it follows that $L_z$ will have a dependence on these values. 

\cite{Donlon2022} (hereafter D+22) used a uniform parallax zero-point offset of $-0.017$ mas to correct the \textit{Gaia} EDR3 parallaxes; in this work, we use the parallax zero-point offset calibration provided by \cite{Lindegren2021}. These different choices of parallax zero-point offsets will result in slightly different calculated distances to each star, which will change the computed $L_z$ value to that star (see \citealt{Helmi2018} for an example of this). 

Similarly, D+22 use a $v_{LSR}$ of 220 km s\invnospace, a Solar position of 8 kpc from the Galactic Center, and a Solar reflex motion of $(10.1, 4.0, 6.7)$ km s\invnospace. This work uses a $v_{LSR}$ of 233.1 km s\invnospace, a Solar position of 8.21 kpc from the Galactic Center, and a Solar reflex motion of $(1.1, 15.17, 7.25)$ km s\invnospace. These choices of kinematic frame will also generate differences in the measured $L_z$ values of each star. 

\begin{figure}
    \centering
    \includegraphics[width=\linewidth]{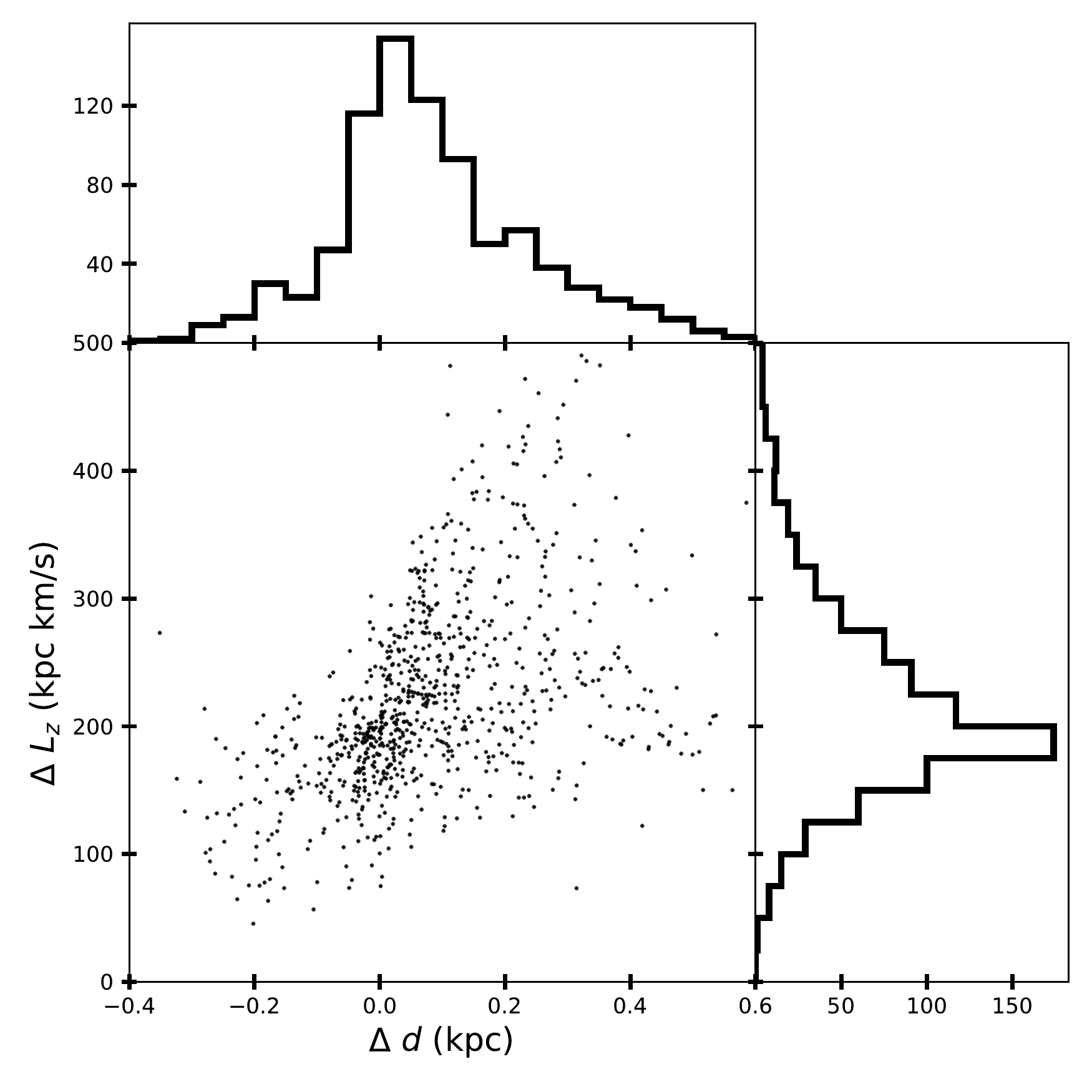}
    \caption{Differences in measured $L_z$ and distance of D+22 and this work, computed for the APOGEE stars in this paper. There is a general correlation between a larger $\Delta d$ and larger $\Delta L_z$. The offset of $\Delta L_z\sim$200 kpc km s\inv at $\Delta d=0$ kpc is due to differences in the phase space coordinates of the Sun.}
    \label{fig:lz_diff}
\end{figure}

It is helpful to determine the impact of this choice of frame on measured values of $L_z$ in order to see how the results of this work compare to the results of D+22. To this end, we define the change in angular momentum,\begin{equation}
    \Delta L_z = L_{z,\textrm{D+22}} - L_{z,\textrm{This Work}},
\end{equation} as well as the change in measured distance,\begin{equation}
    \Delta d = d_{\textrm{D+22}} - d_{\textrm{This Work}}.
\end{equation} Figure \ref{fig:lz_diff} shows the relationship between these values. In general, the larger $\Delta d$, the larger the corresponding $\Delta L_z$. Note the overall offset of $\Delta L_z \sim 200$ kpc km s\inv at $\Delta d=0$: this is because the change in the chosen Solar phase space coordinates produces an offset in the measured $L_z$. 

\begin{figure}
    \centering
    \includegraphics[width=\linewidth]{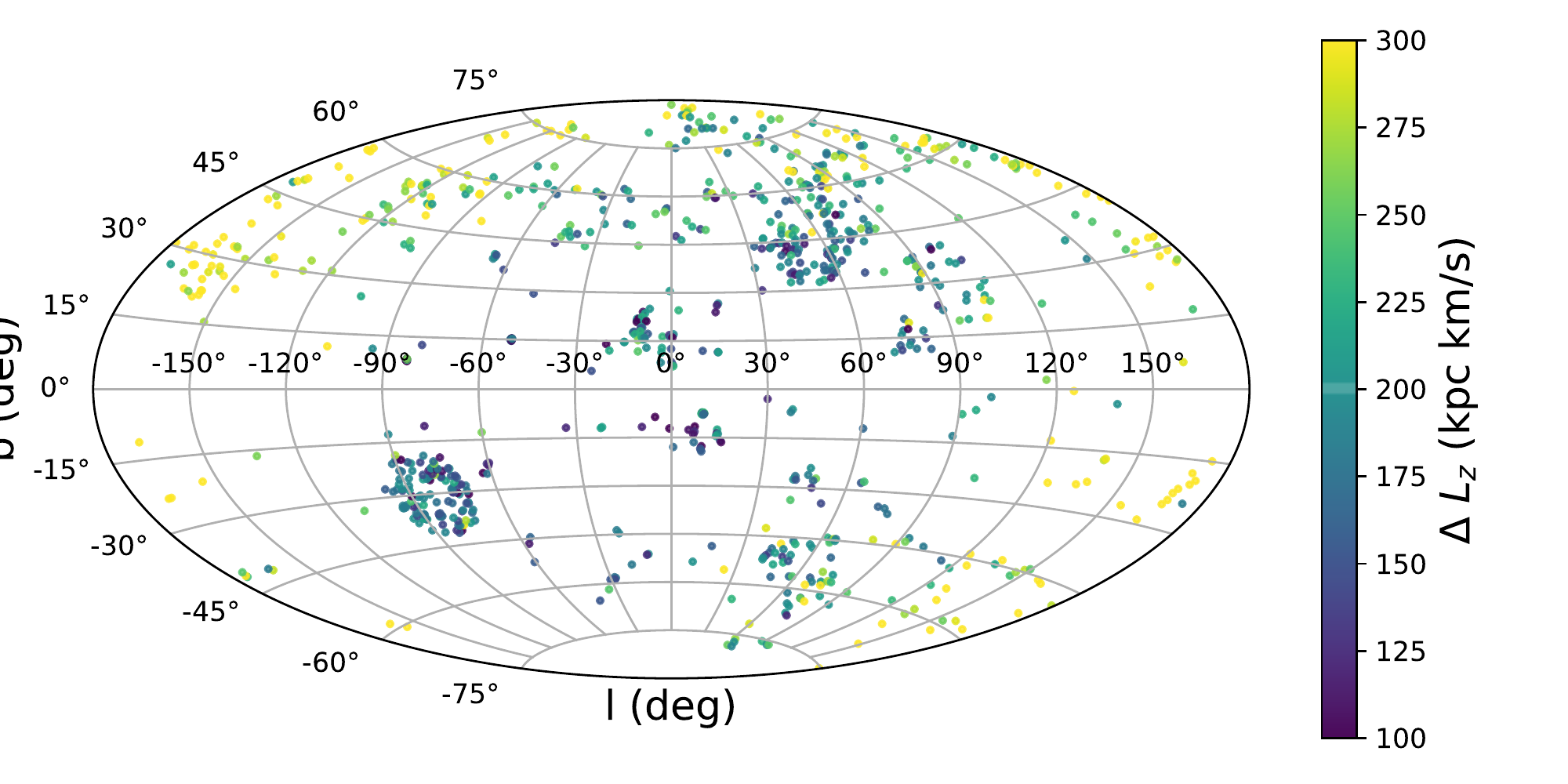}
    \caption{Differences in measured $L_z$ on the sky for APOGEE stars in this work. In general, the closer that a star is to the Galactic Center, the smaller $\Delta L_z$. The stars in D+22 all had $|b|>70^\circ$, which means that their measured $L_z$ values are noticeably different than the stars in this work. }
    \label{fig:lz_diff_sky}
\end{figure}

The variation in $L_z$ is also dependent on other parameters, such as position on the sky. This is shown in Figure \ref{fig:lz_diff_sky}, where it is clear that the measured change in $L_z$ is smallest near the Galactic Center, and is largest in the regions near the Galactic poles and the anticenter. $\Delta L_z$ also varies based on the apparent magnitude of each star, because the parallax zero-point calibration has a complex dependence on the apparent magnitude of each object \citep[Fig. 3 of ][]{Lindegren2021}. 

\begin{figure}
    \centering
    \includegraphics[width=\linewidth]{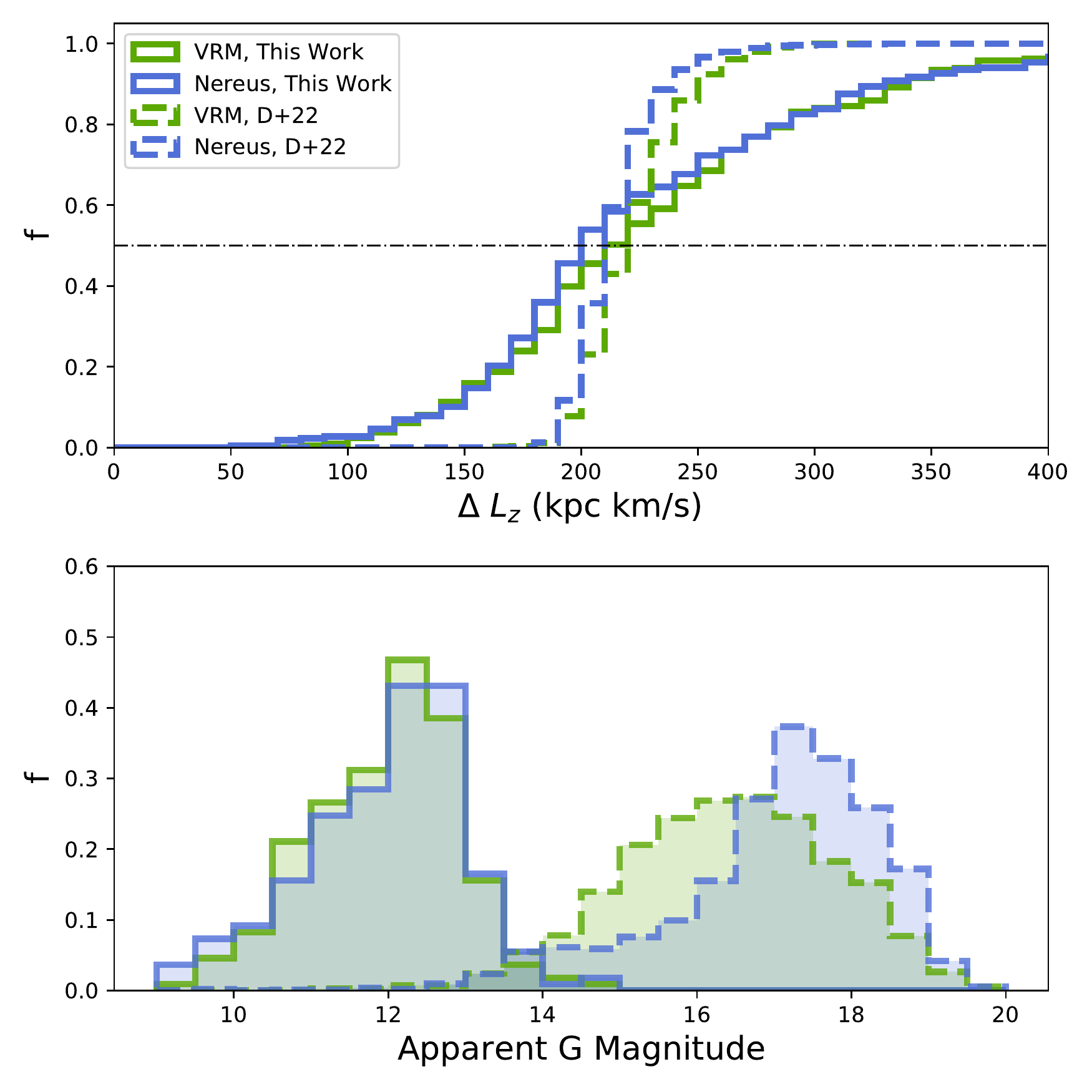}
    \caption{Changes in $L_z$ and apparent magnitude distributions for VRM (green) and Nereus (blue) stars in this work and D+22. Solid lines show the distributions from stars in this work, and dashed lines show the distributions of stars from D+22. \textit{Top:} Cumulative distribution functions of $\Delta L_z$ for VRM and Nereus stars. In both this work and D+22, the mean $\Delta L_z$ is roughly 200 kpc km s\invnospace, as shown by the intersection with the black dashed line at 50\%. The slope of $\Delta L_z$ for the stars in this work is shallower than that of the D+22 stars, because the stars in this work extend out to further distances, which means that $\Delta d$ has a chance to be larger than in D+22 stars. On average, Nereus stars have slightly smaller $\Delta L_z$ than VRM stars. \textit{Bottom:} Apparent magnitude distributions. On average, Nereus stars appear dimmer than VRM stars, which means that they have smaller distances and therefore smaller $\Delta d$, which corresponds to smaller $\Delta L_z$ values than the VRM. However, these offsets are not enough to explain the measured $L_z$ differences in D+22.}
    \label{fig:lz_diff_cdf}
\end{figure}

D+22 found that the VRM had a net $L_z$ of $\sim 400$ kpc km s\invnospace, whereas Nereus had a net $L_z$ of $\sim0$ kpc km s\invnospace. It is important to make sure that this difference in $L_z$ is physical, and not simply due to the choice of kinematic frame and parallax zero-point offset. Figure \ref{fig:lz_diff_cdf} shows the cumulative distribution functions of $\Delta L_z$ split up for VRM and Nereus stars. There is a slightly smaller average $\Delta L_z$ value for Nereus stars, i.e. the $L_z$ values of Nereus stars in D+22 were too small compared to the $L_z$ values of VRM stars, when comparing to $L_z$ values in the frame used in this work. This is likely due to the fact that Nereus stars are on average dimmer than VRM stars (shown in the bottom panel of Figure \ref{fig:lz_diff_cdf}), so Nereus stars will have smaller distances (and also smaller $\Delta d$) on average compared to VRM stars, and therefore a smaller $\Delta L_Z$ as per Figure \ref{fig:lz_diff}. However, this effect is small (only on the order of 10 kpc km s\invnospace), and cannot explain the $\sim$400 kpc km s\inv difference between the two populations measured in D+22. 

We conclude that the choice of frame does have a sizeable effect on measured $L_Z$ values, but it is not able to explain the differences in $L_z$ values of VRM and Nereus stars observed in D+22. The variations between the measured $L_z$ of the VRM and Nereus in this work and D+22 could be due to multiple factors: there is a large dispersion in the measured $L_z$ of these structures in this work, which could ``wash out'' differences between the two substructures; this work primarily studies giant stars, whereas D+22 studies dwarf stars, and the measured $L_z$ of the two populations could be different in some way; or the actual $L_z$ distribution of the merger debris for Nereus and the VRM varies sizeably on the sky, and the regions of the sky that are probed by this work are different enough from those analyzed in D+22 that there is a measured difference in the $L_z$ of each merger event. 

\section{Mock Fit Using APOGEE Abundances} \label{app:mock_fit}

In order to evaluate whether our GMM algorithm is able to recover individual dwarf galaxies as components, we fit our model on a dataset in which it is known to which dwarf galaxy each star belongs. 

For this exercise we used the data from \cite{Hasselquist2021} (hereafter H+21), by matching their Table 2 to the APOGEE DR17 \verb!allStar!catalog. Kinematic parameters were calculated using the APOGEE line-of-sight velocities and the \textit{Gaia} EDR3 proper motions and parallax values provided in the catalog. This yielded a dataset with 6D kinematics and APOGEE chemical abundances for 8005 stars, where H+21 had assigned each star to one of the following structures: the LMC, the SMC, the Sgr dSph plus its tidal tails, Fornax (Fnx), or the GSE. 

\begin{figure}
    \centering
    \includegraphics[width=\linewidth]{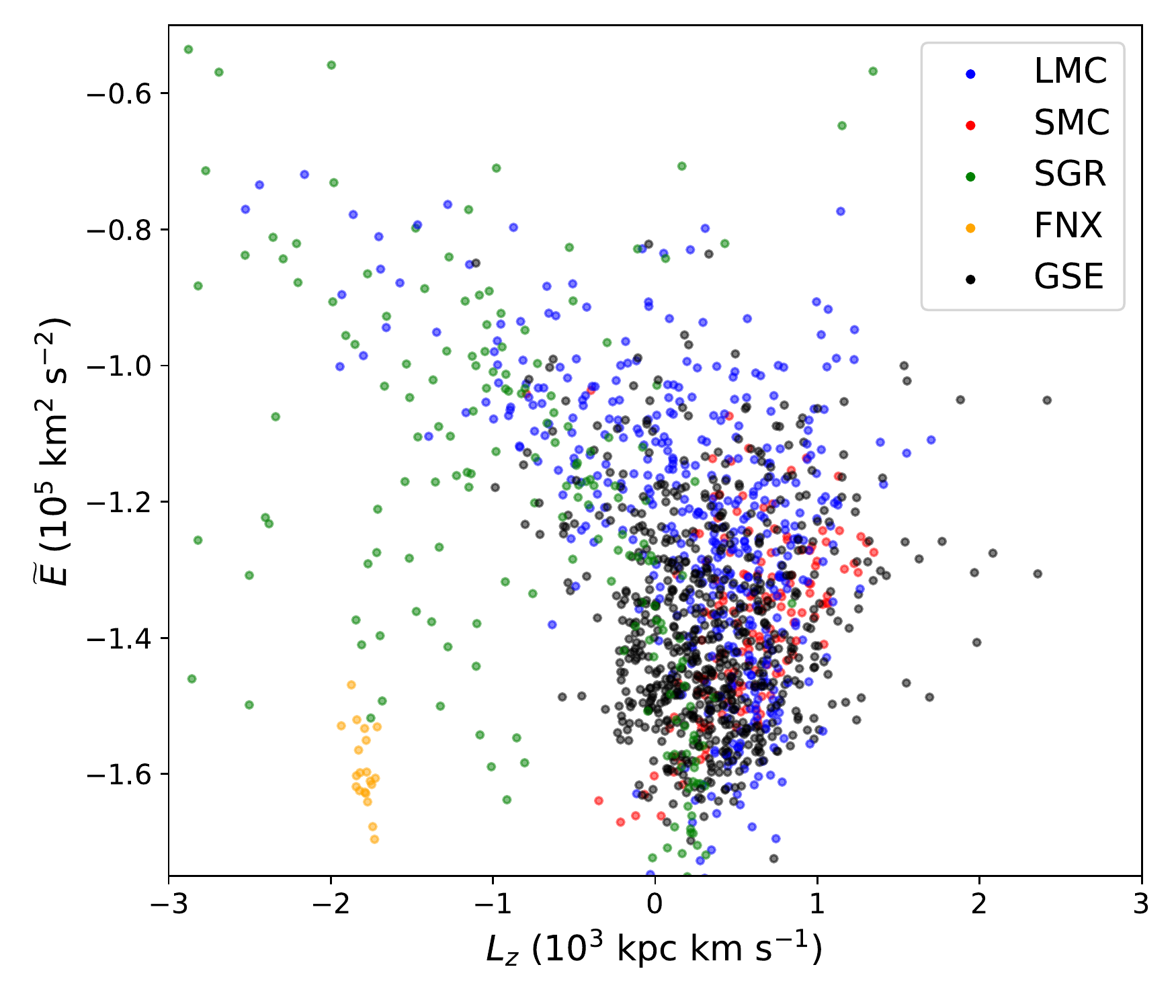}
    \caption{Kinematic quantities for H+21 data. Only stars with \textit{Gaia} DR3 parallax uncertainties less than 100\% are shown (less than 15\% for GSE stars). The stars are colored according to which structure they belong. The LMC, SMC, and Sgr distributions are strongly non-Gaussian; this is likely due to large distance uncertainties in these stars, which severely impacts their measured $\widetilde{E}$ and $L_z$ values. }
    \label{fig:hasselquist_data}
\end{figure}

Figure \ref{fig:hasselquist_data} shows the pseudo-energy $\widetilde{E}$ vs. $L_z$ for stars in the dataset with positive parallaxes and relative parallax uncertainties smaller than 100\%. We only show GSE stars with relative parallax uncertainties below 15\%, because GSE stars can be found closer to the Sun than the other structures in the data. The distributions of kinematic quantities for the LMC, SMC, and Sgr stars are strongly non-Gaussian. This is probably because there are substantial distance uncertainties for most stars; less than 10\% of stars in the data have relative parallax uncertainties below 15\%, and large uncertainties in distance will cause corresponding uncertainties in the measured values of $\widetilde{E}$ and $L_z$. It is difficult to determine more accurate distances to these stars than those calculated from their \textit{Gaia} parallaxes, because they do not share a common spectral type and they do not belong to a single population.

Our GMM algorithm only works well for quantities that are not correlated with one another, which does not appear to be the case for $\widetilde{E}$ and $L_z$ of the H+21 data due to their large distance uncertainties. However, the measured chemical abundances of the APOGEE data are of high quality. In order to utilize the chemical abundances of these stars, we generated a mock set of kinematic data that is guaranteed to have no correlation between $\widetilde{E}$ and $L_z$. We assume that if we had accurate kinematic information for all of the stars in the dataset, the $\widetilde{E}$-$L_z$ distributions of a given dwarf galaxy would be roughly Gaussian. Each structure in the H+21 data was assigned a 2-dimensional normal distribution of $\widetilde{E}$ and $L_z$, where the mean values and variances were chosen to be roughly analagous to those shown in Figure \ref{fig:hasselquist_data}. Then, each star in the dataset was given random $\widetilde{E}$ and $L_z$ values; these values are pulled from the distribution of the structure to which that star is assigned. 

The GMM algorithm was then run on the mock kinematic values for the data, along with their measured chemical abundances. We followed the procedure of Section \ref{sec:model_fit}, i.e. we used the [Fe/H], \fe{\al}, \fe{Na}, and \fe{Al} abundances of the data to fit the model. The best fit model contained 6 components: 1 component corresponded to each of the H+21 structures, plus an additional ``background'' component with a large dispersion in kinematic quantities and chemical abundances, but a small number of stars (less than 5\% of the stars in the dataset). Then, following Section \ref{sec:model_fit}, each star was assigned to a GMM component. 

\begin{figure}
    \centering
    \includegraphics[width=\linewidth]{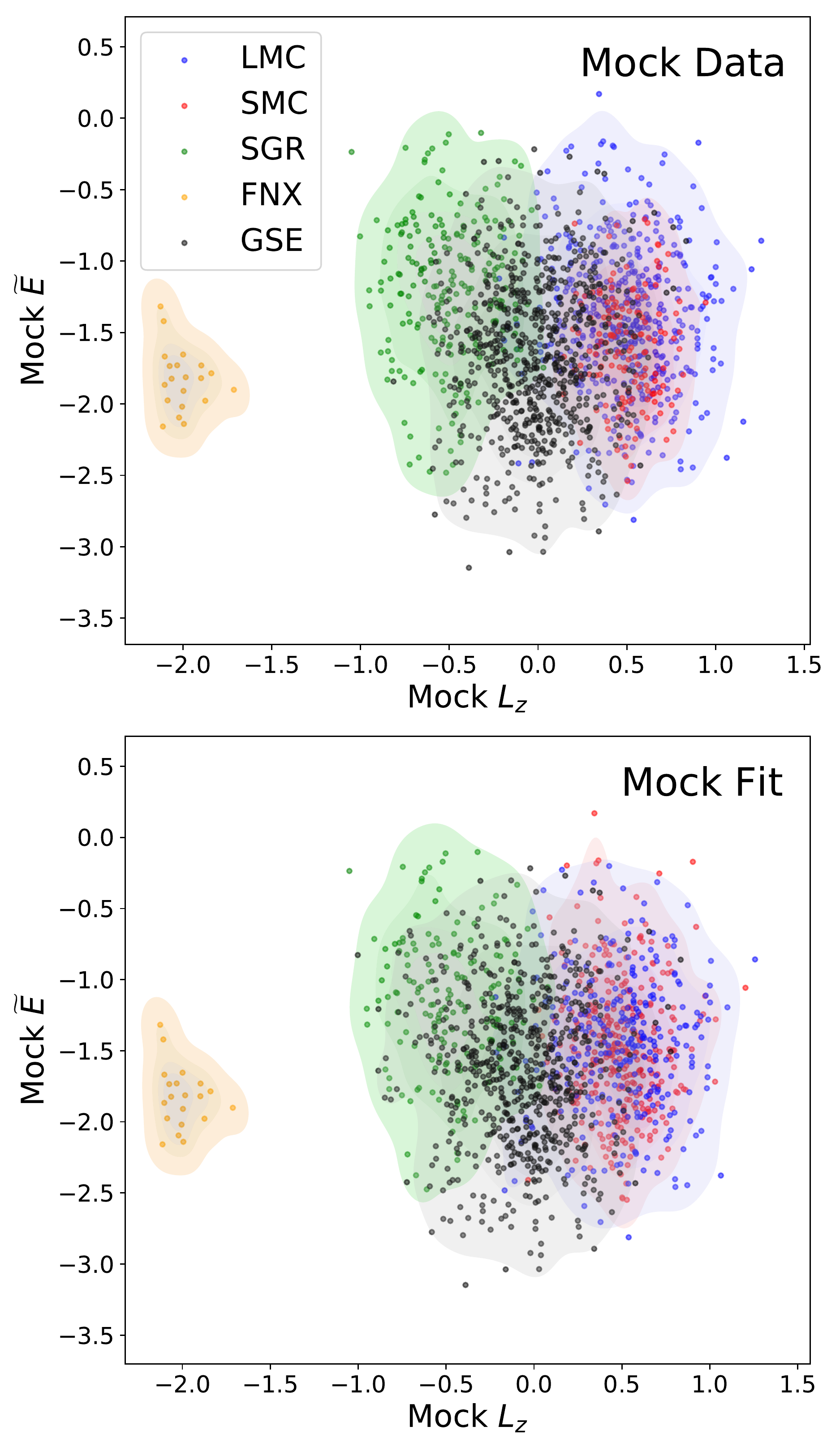}
    \caption{Mock dynamical quantities for the H+21 data (top panel) and the best GMM model fit to that data (bottom panel). The color of each particle corresponds to its structure, or the structure that the model fit assigned to that star. Each structure was assigned a 2-dimensional normal distribution for $\widetilde{E}$ and $L_z$, which were designed to roughly match the observed data in Figure \ref{fig:hasselquist_data}. Then, each star was assigned random $\widetilde{E}$ and $L_z$ values, which were drawn from its component's $\widetilde{E}$-$L_z$ distribution. The mock dynamical values, along with each star's actual chemical abundances, were fed into the GMM algorithm. The best fit GMM algorithm contained 6 components; one for each halo structure, plus an additional ``background'' component (not shown here); the background component has a large dispersion of dynamical and chemical quantities, but a small number of stars (less than 5\% of the total number of stars).}
    \label{fig:hasselquist_fit}
\end{figure}

Figure \ref{fig:hasselquist_fit} shows the mock kinematic distributions of the structures in H+21, as well as the mock kinematic distributions of the fit GMM components. The GMM algorithm does a good job of recovering the underlying components of the distribution; however, this may not be particularly surprising or interesting considering a GMM was used to generate the mock kinematic data. The main difference between the true distributions and the fit distributions is that the algorithm incorrectly assigned several of the LMC stars to the SMC distribution. 

\begin{figure*}
    \centering
    \includegraphics[width=\linewidth]{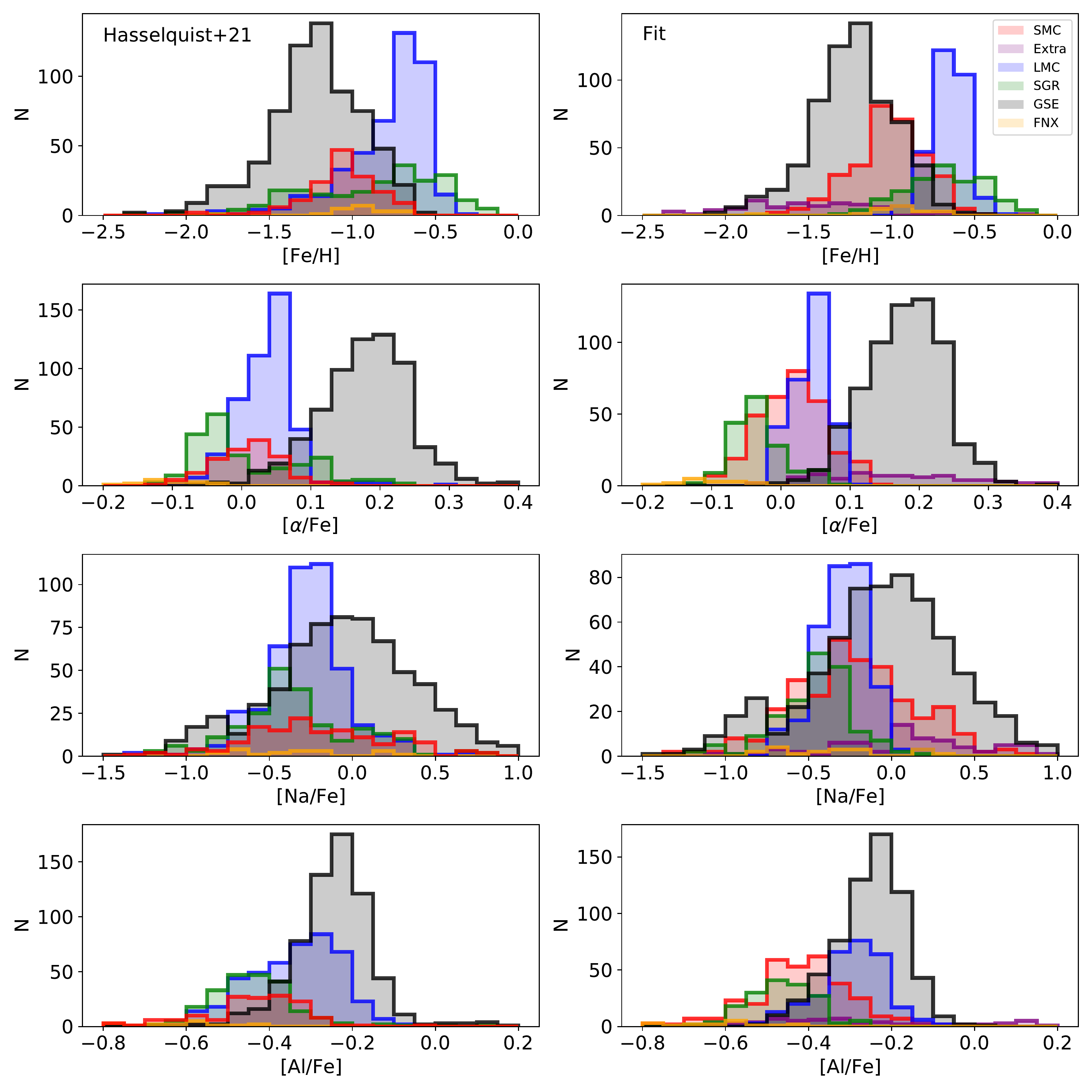}
    \caption{Chemical abundance distributions for the H+21 data (left column) and the model fit (right column). Stars from each structure are split up by color. Overall, the fit does a good job of reproducing the input distributions. However, the best model fit placed many stars in the SMC distribution that really belonged to the LMC, and modeled an extra ``background'' component (purple) with a small number of stars. The model also struggled to associate a small number of [Fe/H]-poor, \fe{\al}-rich stars in Sgr with the Sgr component (green); this may be because the Sgr component appears to have multiple peaks in [Fe/H] and \fe{\al}.}
    \label{fig:hasselquist_cdfs}
\end{figure*}

Figure \ref{fig:hasselquist_cdfs} shows the chemical distributions of the H+21 structures compared to the fit distributions. Overall, the fit model does a good job of recovering the true distributions of the data. As seen in the mock kinematic data, a number of stars that actually belong to the LMC are assigned to the SMC by the model. The model also struggled to associate a small number of [Fe/H]-poor, \fe{\al}-rich stars in Sgr with the Sgr component (green); this may be because the Sgr component appears to have multiple peaks in [Fe/H] and \fe{\al}. 

We also ran the GMM algorithm on the H+21 data using the measured kinematic quantities, in order to explore how robust the model is for non-Gaussian distributions. The best fit model had the correct number of components, and it was able to assign 68\% of the stars to the correct components, compared to the mock kinematics, where the algorithm was able to correctly assign 78\% of stars to the correct component. It is possible that some stars in the H+21 data do not actually belong to the dwarf galaxy to which they were assigned, which would reduce our algorithm's ability to correctly identify to which component a given star belongs.

We are confident that the GMM algorithm that we present in this work is able to recover distributions of stars that belong to one or more dwarf galaxies, with fairly high accuracy, even when those distributions are complicated and overlapping. 

\end{document}